\documentclass[a4paper,10pt]{article}
\usepackage{a4wide,makeidx}                             
\usepackage[english]{babel}                             
\usepackage{amsmath,amsfonts,verbatim,graphicx,float} 

\begin{document}

\begin{center}
{\Large \bf Baryons from quarks in curved space
 and deconfinement\/}
\end{center}

\vspace{0.5cm}
\begin{center}
M.\ Kirchbach and C.\ B.\ Compean,\\
Instituto de Fis{\'{\i}}ca,
Universidad Autonoma de San Luis Potos{\'{\i}},\\
Av. Manuel Nava 6, Zona Universitaria,\\
S.L.P. 78290, M\'exico
\end{center}

\vspace{0.5cm}
\begin{flushleft}
{\bf Abstract:}
Detailed account is given of the fact that
the Cornell potential predicted by Lattice QCD
and its exactly solvable 
trigonometric extension recently reported by us 
can be viewed as the respective approximate and exact counterparts
on a curved space to  an $1/r$ flat space potential.
 The ``curved'' potential describes a confinement phenomenon as it
is of infinite depth and has only bound states.
It furthermore has the remarkable property of preserving both 
the $SO(4)$ and $SO(2,1)$
symmetries characterizing the ordinary $1/r$  potential.
We first make the case that this particular geometric vision on confinement
provides a remarkably  adequate description of both nucleon and $\Delta $
spectra and the proton mean square charge radius as well, and
suggests an intriguing venue toward quark deconfinement as a shut-down of 
the curvature considered as temperature dependent.
Next we observe that the $SO(2,1)$ symmetry of the ``curved'' potential
allows to place it within  the context of  
$AdS_5/CFT$ correspondence and to establish in this manner the 
algebraic link of the latter to QCD potentiology. 

\end{flushleft}

\vspace{0.5cm}

\begin{flushright}
``...there will be no contradiction in our mind if we assume that some natural
forces are governed by one special geometry, while other forces by another.''
\end{flushright}

\vspace{0.05cm}

\begin{flushright}
N.\ I.\ Lobachevsky
\end{flushright}
\section{Introduction}
One of the major achievements  of contemporary 
physics concerns the insight on the non-trivial geometry of the Universe. 
According to general theory of relativity,
the space is curved by the presence of mass, a well established
concept which was successful in explaining the precession of the 
orbit of Mercury and the bending of light in the vicinity of the Sun.
Various geometries have been under consideration ever since in gravity,
the closed Einstein's universe of constant positive  
and the open one of Lobachevsky of constant negative curvature  
being among the most prominent  versions
(see \cite{Gersten},\cite{Carin} for contemporary treatises). 
On long term, the ideas of general relativity exercised a profound 
impact on the development of quantum physics.
On the one hand, they led to the 
geometric interpretation of gauge theories 
(see \cite{gauge} for a pedagogic presentation)
and on the other, they triggered  progress
in merging  external space-time
with  internal gauge degrees of freedom 
which culminated in superstring and supergravity theories.

Perhaps nothing expresses relevance of the above movements better but the
surprisingly modernly sounding statement made by  Lobachevsky
approximately 200 years ago (taken from ref.~\cite{Vozmi})
which we used  as motto to the present article.
Indeed, the geometric view on both space time and
gauge theories suggests that some of the fundamental 
physics phenomena might be related to curvature,
an obvious candidate being the color confinement phenomenon.

The idea that quark confinement might reflect some kind of curvature  
has been pioneered by Salam and Strathdee \cite{SS_78} who found a
black-hole solution of Einstein's equation approximate to the
anti-de Sitter $AdS_5$ geometry which  describes strongly
interacting tensor fields confined in a micro-universe of a radius 
fixed by the negative cosmological constant and which they interpreted as
a sort of hadron bag. 
In this manner, particles have been described within 
gravitational context as strongly curved universes,
an idea pursued by several authors within various contexts \cite{bholes}.  
Among the achievements of this idea we count
(i) the explanation of the flavor independent level spacings
of the radial excitation spectra of mesons following from the
local isomorphism between the anti-de Sitter group $SO(3,2)$ 
and $Sp(4,R)$, the symmetry group of the two-dimensional harmonic oscillator,
(ii)  the understanding of quark confinement
as divergence of the bag radius due to its thermal dependence 
leading to space flattening \cite{Takagi}.
Some of the constituent quark models, 
such as the MIT bag, the  ``Chashire cat'' 
or the Skyrme models have been interpreted (roughly speaking) 
as black holes of anti-de Sitter geometry.      

From the outgoing 90ies onward, 
the idea of the geometric confinement and the study of AdS space-time
manifolds experienced a strong push, now from the new perspective of the 
$AdS_5/CFT$ correspondence according to which a
maximal supersymmetric Yang-Mills conformal field  theory (CFT)
in four dimensional Minkowski space is equivalent to a 
type IIB closed superstring theory 
in ten dimensions described by the product manifold
$AdS_5\times S^5$ \cite{AdS}.
More recently,  the correspondence between 
string theory in ten dimensional anti-de Sitter 
space and $SO(4,2)$ invariant conformal Yang-Mills theories 
has been adapted in \cite{Brodsky} to the description
of hadron properties.

Parallel to the above achievements, significant progress in 
understanding  hadron properties has been reached    
independently through the elaboration of the connection between
the QCD Lagrangian and the potential models 
as deduced within the framework of effective field theories
and especially through the  non-perturbative methods such as 
lattice simulations \cite{Bali},\cite{Brambilla}, the most prominent 
outcome being the linear plus Coulomb confinement potential \cite{Lattice},
\cite{Cornell}. The potentials derived from the QCD Lagrangian
have been most successful in the description
of heavy quarkonia and heavy baryon properties \cite{Nora_report}. 
Although the Cornell potential has found applications also in 
nucleon and $\Delta $ quark models \cite{varca}, 
the provided level of quality in the description in the non-strange sector 
stays  below the one reached for the heavy flavor sector.
This behavior reflects  insufficiency of the one gluon exchange
(giving rise to  the Coulomb-like term) and of the flux-tube interaction
(associated with the linear part) to account for the complexity
of the dynamics of three light quarks.
Various improvements have been under consideration 
in the literature such as screening effects in combination with spin-spin
forces (see \cite{garz} and reference's therein).

Very recently, the Cornell potential has been updated through 
its extension toward an exactly solvable 
(in the sense of the Schr\"odinger equation, or, 
the Klein-Gordon equation with equal scalar and vector potentials)
trigonometric quark confinement potential \cite{Quiry_2}.
The latter potential, which has the property of
interpolating  between a Coulomb-like potential
and the infinite well while passing through a region of linear growth,
was shown to provide a remarkably good description of
the spectra of the non-strange baryons, the nucleon and the $\Delta$,
considered as quark-diquark systems, 
and of the proton charge radius as well. Especially the observed
hydrogen-like degeneracy patterns in the above spectra
in the non-trivial combination with the non-hydrogen like (because increasing 
instead of decreasing) level spacings found a stringent explanation
in terms of the $SO(4)$ symmetry of the trigonometric potential.
   
\begin{quote}
The main goal of the present work is to draw attention to the fact that
similarly to the Coulomb-potential, the trigonometrically extended 
Cornell (TEC) potential,  possesses next to $SO(4)$ also $SO(2,1)$ symmetry
as manifest through the possibility to cast its spectrum as well in terms 
of  $SO(4)$ as in terms of  $SO(2,1)$ Casimirs.
The $SO(2,1)$ versus $SO(4)$ symmetry correspondence allows to place the 
TEC potential within the context of
 $ AdS_5$ versus $CFT$ correspondence and in 
this manner to link the algebraic aspects of $AdS_5/CFT$  
to QCD potentiology \cite{Bali}. 
Such is possible because 
while $SO(2,1)$ appears in one of the possible $AdS_5$ 
reduction chains \cite{Fischer}, namely,
\begin{equation}
SO(3,2)\subset SO(2,2)\subset \underline{SO(2,1)}\subset SO(2),
\label{AdS_groupchain}
\end{equation} 
$SO(4)$ appears within the  reduction chain associated with $CFT$,
 \begin{equation}
SO(4,2)\subset SO(4,1)\subset \underline{SO(4)}\subset SO(3)\subset SO(2).
\label{CFT_groupchain}
\end{equation} 

\end{quote}
It is important to be aware of the fact that the algebraic
$AdS_5/CFT$ criteria alone
are not sufficient to fix uniquely the potential.
One  has to complement them by the requirement on compatibility with 
the QCD Lagrangian too, a condition which
imposes severe restrictions on the allowed potential shapes.
\begin{quote}
Our next point is that algebraically 
the $AdS_5/CFT$ correspondence translates into
$SO(2,1)/SO(4)$ symmetry correspondence of the potential and that it is
the trigonometrically extended Cornell potential, 
treated as quark-diquark potential,
the one that meets best both the $AdS_5/CFT$ and QCD criteria
and provides the link between them.
Predicted degeneracy patterns and level splitting are such
that none of the observed  $N$ states 
drops out of the corresponding systematics
which also applies equally well to the $\Delta $ 
spectra (except accommodation of the hybrid $\Delta (1600)$). 
The scenario provides
a remarkable description of the  proton charge electric
form-factor too and moreover implies a deconfinement mechanism
as a shut-down of the curvature considered as temperature dependent.
\end{quote}

The group symmetries under discussion appear within the context of
a Schr\"odinger equation written in different variables. 
Specifically for the case
of the hydrogen atom these different variables have been extensively studied 
and are well known. The $SO(4)$ appears as symmetry of
the standard radial part, $R(r)$, 
of the Schr\"odinger wave function,
with $r$ standing as usual for the radial distance, while
$SO(2,1)$ is the symmetry of same equation when transformed to
$r=y^2$ and $R(r)=y^{-\frac{3}{2}}Y(y)$ variables 
\cite{Wybourne},\cite{Marcelo}.
Obviously, in the specific case under consideration,
both symmetries are physically indistinguishable
because they both lead to same physical observables such as
spectrum and transition probabilities. Preferring the one over the other
as mathematical tool in hydrogen description 
is a pure matter of convenience and  giving preference to  $SO(4)$ 
is only more popular. The Coulomb potential is 
an example that matches algebraically $AdS_5/CFT$ correspondence 
but is at odds with the non-perturbative aspects of QCD dynamics.

The TEC case of major interest in this work 
presents itself bit more involved. While the Schr\"odinger equation giving rise
to the $SO(4)$ symmetric spectrum is well studied and well 
understood in terms of  a potential satisfying the 
Laplace-Beltrami equation on the three dimensional (3D) 
hypersphere, $S^3_R$, of constant radius, $R$, i.e. on
a curved space of a constant positive curvature, 
knowledge on the differential realization of the $SO(2,1)$ 
symmetry on a hyperbolic space
is quite scarce indeed (see next section).  Filling this 
technical gap should certainly be beyond the scope of the present study 
which main focus are spectroscopic observables. However, a strong
though indirect hint on the relevance of $SO(2,1)$ for the TEC problem is 
provided by the possibility to recast its spectrum in terms of 
$SO(2,1)$ Casimir eigenvalues (admittedly,  for limited values of the
strength parameter \cite{Koca}). In other words, at least
for some particular values of the strength parameters  a  manifest coordinate 
transformation of the Schr\"odinger equation with the TEC potential 
from a four-dimensional Euclidean to a three dimensional hyperbolic 
geometry is likely to exist. Of course, unless
such a transformation has not been explicitly constructed,
no algebraic formalization of a possible indistinguishability between 
the respective $SO(4)$ and $SO(2,1)$ symmetries can be claimed.
Rather, one has to make a side by side comparison of the predictions
of  both schemes with the aim of proving their proximity, 
a subject treated in Section 4 below.

Finally, besides viewing $SO(4)$ as part of $CFT$, it can also be
embedded within an (admittedly, Euclidean) anti-de Sitter space
\cite{Maximo}, 
\begin{equation}
-x_5^2 + x_0^2 +x_1^2 +x_2^2+x_3^2 =-l^2,
\label{AdS_Eucl}
\end{equation}
where the $x_i$ with $i=1,2,3$ are the usual Cartesian coordinates
in standard position space, $x_5$ and $x_0$ are the 
two time-like dimensions,  from which the second has been ``Wick rotated'',
and $-1/l^2$ is the negative curvature.
This additional insight provides, in our view, a 
further legitimization for accepting, 
without too much a loss of generality, the hypersphere
$x_0^2+x_1^2+x_2^2+x_3^2=R^2$ as a mathematical tool
in the description of quark confinement as infinite potential barrier,  
a venue that takes directly to the trigonometric 
extension of the Cornell potential. In taking this path, however, 
one should treat the positive curvature, 
$\kappa=1/R^2$, introduced in that manner, with some care
and  detain from equipping it with too deep a physical meaning. Rather
it should be viewed as a second phenomenological parameter next to 
the potential strength which so far stays uncorrelated  to the physical 
cosmological constant, $\lambda =-1/l^2$. 
We shall show that while spectra
and charge form factors remain by and  large insensitive to this parameter,
it provides a valuable phenomenological 
tool for deconfinement description in the spirit of 
ref.~\cite{Takagi}.

The outline of the  paper is as follows.
The next section is a historic survey on the quantum Kepler problem
in a space of constant positive curvature, the 3D hypersphere,
$S_R^3$, a survey that begins with early work by Schr\"odinger ~\cite{Schr40}. 
A detailed account is given of the fact that
the harmonic potential on $S_R^3$, i.e. the one that
satisfies the four-dimensional angular
Laplace-Beltrami equation, takes the form of
the trigonometric confinement potential 
$-2b\cot \chi +l(l+1)\csc^2 \chi $ (with $\chi $ standing for the 
second polar angle). Depending on the $\chi $ parametrization in terms
of coordinates in ordinary position space, a variety of 
potentials can be created.  
Examples are the  
the exactly solvable trigonometric extension of the Cornell quark 
confinement potential, 
around which the present work is centered, and which corresponds to
$\chi =r\sqrt{\kappa} \pi $  where $r$ is the absolute value of 
the radius vector, and $\kappa $ is the curvature of the hypersphere.
Another version would be a   
gradient dependent confinement potential
for particles with position and curvature dependent mass as needed 
for the purposes of quantum dots. We attend  also this more subtle
version because of its possible relevance for the description of the
evolution of finite valence to vanishing parton quark masses.
These two  examples are presented in section 3.
Section 4 focuses on the application of 
the trigonometric extension of the Cornell potential
to  the spectra of the nucleon and the  $\Delta $ considered as
quark--diquark systems. It further contains the description of same spectra
within the $SO(2,1)$ symmetric version of same potential thus 
revealing its link to the algebraic aspects of the $AdS_5/CFT$ scenario. 
The section ends with a calculation of
the mean square proton charge radius. 
Section 5 presents the property of
energy spectrum and  wave functions
of the TEC problem to collapse 
upon curvature shut-down (i.e. in the  large $R$ limit) 
to the bound and scattering states of  ordinary 
flat-space $1/r$ potential, a peculiarity that we employ
(in parallel to Takagi's work \cite{Takagi} mentioned above)
to interpret  deconfinement as flattening of space due to 
a thermal  dependence of the curvature parameter.
The paper closes with  brief summary and outlooks.

\section{Harmonic  potential on $S^3_R$:The survey }

The first to have  considered the Coulomb potential on a curved space 
has been Schr\"odinger who solved in
 \cite{Schr40} the quantum mechanical Coulomb problem in the 
cosmological context of Einstein's universe, i.e. 
on the three dimensional  (3D) hypersphere, 
$S_R^3$,  of a constant radius $R$.
Schr\"odinger's prime result, the presence of curvature provokes 
that the orbiting particle appears confined 
within a  trigonometric potential of infinite depth
and the hydrogen spectrum shows only bound states. 
An especially interesting observation
was that the $O(4)$ degeneracy of the levels observed in the 
flat space $H$ atom spectrum was preserved by
the curved space spectrum too in the sense that
also there the levels could be labeled by the standard atomic indices
$n$, $l$, and $m$, and the energy depended on $n$ alone. 
However, contrary to flat space, no explicit form of $O(4)$
 generators could be immediately exhibited. Although
Higgs \cite{Higgs} and Leemon \cite{Leemon} succeeded in constructing on 
$S^3_R$  the respective  analogue to the Runge-Lenz vector in flat space,
no way was found to incorporate it into the $O(4)$ group algebra.
Instead, Barut and collaborators \cite{Barut} designed a version of
the potential as a differential $su(1,1)$ Casimir operator in 
allowing one of the potential parameters to
depend on the principal quantum number in a very particular way.  
However, this version, strictly speaking, does not share the 
original $SO(4)$ degeneracy patterns and is not of 
interest to the present study.

Perhaps because Schr\"odinger used the curved space Coulomb problem
as an example
for exact solubility of his celebrated equation 
by means of the factorization technique, it became more popular
in that very context than in any other giving rise to
the field of physics known as \underline{su}per\underline{sy}mmetric 
\underline{q}uantum \underline{m}echancs (SUSYQM),
a historical development mainly triggered by subsequent extensive
work by Infeld and collaborators \cite{Infeld} and later on by Witten 
\cite{Witten_S}. 

Nonetheless, also the geometric aspect of 
Schr\"odinger's work was independently picked up by several researchers 
and placed within various contexts.
 The idea of using such  ``curved'' potentials
gradually breached into  several areas of 
quantum physics from
atomic \cite{bessis} to the utmost modern 
nano-tubes physics \cite{kourochkin},  
 and
sophisticated  non-linear $W$ algebra symmetries \cite{boer}
viable in string theories.  Bessis et al. \cite{bessis} 
applied it to fine structure analysis of atomic spectra,
Ballesteros and Herraz \cite{Balle}
considered it within the context of quantum algebras, while
Roy and Roychoudhuri formulated SUSYQM in a 3D curved space \cite{Pinaki}. 
Also the Russian school provided notable contributions especially
regarding the mathematical aspects of the solutions \cite{Vinitski}.
 
It is worth noticing that also the harmonic oscillator (HO) potential has been
considered on $S^3_R$ in \cite{Higgs}, \cite{Bonatsos}.
Finally,
both the Coulomb and the HO problems have been also solved in spaces with a 
negative constant curvature (Lobachevsky geometry) in the second reference 
~\cite{Infeld}, and in
\cite{DelSolMesa}, \cite{Barros}, to mention only
few representative examples of such studies (see also ref.~\cite{Carin}
for a recent up-date).

The current section is a historic survey on the quantum Kepler problem
in a space of constant positive curvature, the 3D hypersphere,
$S_R^3$. It contains a detailed account of the fact that
the harmonic potential on $S_R^3$, i.e. the one that
satisfies the four-dimensional Laplace-Beltrami equation, takes the form of
the trigonometric confinement potential 
$-2b\cot \chi +l(l+1)\csc^2 \chi $ (with $\chi $ standing for the 
second polar angle).

\subsection{ $S^3_R$ parametrization and curved space Coulomb-like
 potential }
From now onward the usual three dimensional
flat Euclidean space, $E_3$, will be 
embedded in the four dimensional Euclidean  space, $E_4$.
A set  of generalized Cartesian coordinates, 
$\lbrace x_1,x_2,x_3,x_4 \rbrace $, in $E_4$ chosen to 
parametrize a three-dimensional spherical surface there,
has to satisfy the condition,
\begin{equation}
s^2=x_1^2+x_2^2+x_3^2 +x_4^2, \quad 
\kappa = \frac{1}{s^2 } \, ,\quad 0<s<\infty,
\label{E_4}
\end{equation}
where $s$ is the hyper-radius, 
and  $\kappa $ the corresponding  curvature.
A Coulomb-like potential in any $E_n$ space is
defined from the  requirement  to be harmonic, i.e. to
obey the respective $n$-dimensional Laplace-Beltrami 
equation in charge free spaces. 
Specifically in $E_4$ such a  potential, call it $v(\bar x)$,
where $\bar x$ denotes the radius vector of a generic  point on 
the hypersphere,  
is most easily found in Cartesian coordinates 
\begin{equation}
\Box v(\bar x)=\sum_{i=1}^{i=4}\frac{\partial^2}{\partial x_i^2} 
v(\bar{x})=0,
\label{Laplace_4}
\end{equation}
and reads
\begin{equation}
v(\bar{x})= c \frac{x_4}{{\bar r}}, \quad \bar r=
\sqrt{x_1^2+x_2^2+x_3^2}.
\label{Coul_4}
\end{equation}
Here, $\bar r$ is the length of the radius vector in the
$E_3$ subspace of $E_4$,   
$c$ is a constant, and use has been made of the fact that
the $1/{\bar r}$ potential is harmonic in $E_3$ as it
satisfies there the three-dimensional  Laplace equation, 
${\vec \nabla}^2 (1/{\bar r})=0$.
Changing now to hyperspherical coordinates,
$\Omega=\lbrace \chi, \theta ,\varphi\rbrace $,
results in 
\begin{eqnarray}
x_1=\bar r \sin\theta \cos\varphi, &\quad &
x_2= \bar r\sin\theta \sin\varphi,\nonumber\\
x_3= \bar r \cos\theta, &\quad & x_4 =s\cos \chi,\nonumber\\
\bar r=s\sin\chi, \quad 0\leq \chi \leq \pi, &\quad &0\leq \theta \leq \pi,
\quad 0\leq \varphi  \leq 2 \pi.
\label{4d_sphr}
\end{eqnarray}
In terms of $\chi $, the ``curved'' $1/{\bar r}$  potential
is read off from eq.~(\ref{4d_sphr}) as the following 
trigonometric potential,
\begin{equation}
v\left(\frac{x_4}{\bar r} \right)\equiv v(\chi   )=
c \cot \chi\, .
\label{harm_4_pot}
\end{equation}
This is a very interesting situation in so far as in $E_4$
the r\'ole of  the radial coordinate of infinite range in 
ordinary flat space, $0<{\bar r}< \infty $, has been taken by the
angular variable, $\chi$, of finite range. In other words, 
while the harmonic potential in $E_3$ is a  central one, in
$E_4$ it is non-central. Moreover, 
\begin{quote}
the inverse distance potential of finite
depth in $E_3$ is converted to an infinite barrier and therefore to a
confinement potential in the higher dimensional $E_4$ space, 
a property  of fundamental importance throughout the paper. 
\end{quote}

\subsection{Schr\"odinger's geometric quantum scheme }
This section contains a detailed account of Schr\"odinger's treatment
\cite{Schr40} of the  quantum mechanical Coulomb problem 
within Einstein's cosmological concept, i.e. on the three dimensional
sphere of constant radius, $S_R^3$. For this purpose, 
Schr\"odinger had to solve his celebrated  equation in $E_4$.
The equation is not only especially simple to solve on the 
hypersphere of constant radius, $s=R=$const, where it is purely angular, 
but it is also there where it acquires a special physical importance, 
to be revealed below.  

When written in the hyper-spherical coordinates, it takes the following form,
\begin{equation}
\left( -\kappa\frac{\hbar ^2}{2\mu } {\widehat \Box} +c  \cot \chi \right)
\Psi (\chi, \theta, \varphi, \kappa)=E (\kappa )
\Psi (\chi, \theta, \varphi, \kappa), \quad \kappa =\frac{1}{R^2}=\mbox{const},
\label{Schroed_4}
\end{equation}
where $\widehat{\Box}$ denotes the angular (hyperspherical) 
part of the $E_4$ Laplace-Beltrami
operator.
Here, and without loss of generality, the potential strength has been kept
unspecified so far and denoted by the constant $c$.
Using the well known representation of $\widehat{\Box}$,
\begin{eqnarray}
 {\widehat \Box} &=&  \left[\frac{1}{\sin^2\chi }
\frac{\partial }{\partial \chi}
\sin^2\chi \frac{\partial }{\partial \chi } -
\frac{L^2 }{\sin^2 \chi }\right],\nonumber\\
L^2&=&-\left[ \frac{1}{\sin\theta }\frac{\partial }{\partial \theta} 
\sin\theta \frac{\partial }{\partial \theta }
+\frac{1}{\sin^2\theta } \frac{\partial^2}{\partial \varphi^2}
\right],
\label{lpls_4}
\end{eqnarray}
where $L^2$ is the standard three dimensional orbital 
angular momentum operator in $E_3$,
and separating variables in  the solution as
$\Psi (\chi, \theta, \varphi)=\psi (\chi ,\kappa )Y_l^m(\theta ,\varphi)$,
the following equation in the $\chi$ variable (hyper-angular equation) 
emerges,
\begin{eqnarray}
{\Big[} -\kappa \frac{\hbar^2}{2\mu } \frac{1}{\sin^2\chi }
\frac{\partial }{\partial \chi }
\left( \sin^2\chi \frac{\partial}{\partial \chi } \right)
&+& {\mathcal V}(\chi ,\kappa )-E(\kappa ){\Big]} \psi  (\chi, \kappa )=0,
\nonumber\\
{\mathcal V}(\chi ,\kappa )&=& 
\kappa \frac{\hbar^2}{2\mu } \frac{l(l+1)}{\sin^2\chi }+ c\cot \chi \, . 
\label{chi_eq}
\end{eqnarray}
Multiplying eq.~(\ref{chi_eq}) by $\left( -\sin^2\chi \right)$ and 
changing variable to
\begin{equation}
X (\chi, \kappa  )=\sin \chi 
\psi (\chi,\kappa ),
\label{trsfm}
\end{equation}
allows to cast it  into the form of the following
one-dimensional Schr\"odinger equation in
the angular  variable $\chi $,
\begin{eqnarray}
\left[ -\kappa \frac{\hbar^2}{2\mu }\frac{\mbox{d}^2  }{\mbox{d}\chi ^2}
+ {\mathcal V}(\chi ,\kappa)\right] X(\chi ,\kappa ) 
&=& \left(E(\kappa )+ \kappa \frac{\hbar^2}{2\mu }
 \right)X (\chi, \kappa ).
\label{chi_world}
\end{eqnarray}
This is precisely the equation first obtained by 
Schr\"odinger \cite{Schr40}. Before proceeding further,  
it is quite instructive to 
first take a close look on the  free particle motion on 
$S_R^3$, i.e. $c=0$, a subject treated in the next section.

\section{Confinement phenomena as infinite barriers due to 
curvature:The scenarios}

\subsection{Confinement as centrifugal barrier of free particle motion 
on $S^3_R$}
{}For a free particle motion on $S^3_R$  equation ({\ref{chi_world})
reduces to 
\begin{equation}
\left[ 
-\kappa \frac{\hbar^2}{2\mu }\frac{\mbox{d}^2 }{\mbox{d}\chi ^2}
+\kappa \frac{\hbar^2}{2\mu }\frac{l(l+1)}{\sin^2\chi }\right]
{\mathcal S} (\chi,\kappa  ) 
= E^{(c=0)}(\kappa ) {\mathcal S} (\chi ,\kappa  ),
\label{chi_free}
\end{equation} 
with ${\mathcal S}(\chi ,\kappa )$ denoting the free-particle solution.
The second term on the l.h.s. of this equation
describes the centrifugal energy, $U_l(\chi, \kappa)$, of a particle of a 
non-zero orbital angular momentum 
on $S^3_R$, i.e.,
\begin{equation}
U_l(\chi ,\kappa) =\kappa \frac{\hbar^2 }{2\mu }\frac{l(l+1)}{\sin^2 \chi }.
\label{4_cntrf}
\end{equation} 
This  term provides an infinite barrier and 
thereby a confinement, an observation to acquire profound 
importance in what follows.

The energy spectrum of eq.~(\ref{chi_free}) is easily found from the
observation on its inherent  $O(4)$ symmetry. Indeed,
the angular part, ${\widehat \Box} $, of the four-dimensional  
Laplacian, $\Box$,  represents the operator of the
four-dimensional angular momentum, 
here denoted by ${\mathcal K}^2$,
\footnote{The analogue on the two-dimensional sphere of a constant radius
$r=a$  is the well known relation
${\vec \nabla}^2 =-\frac{1}{a^2}L^2 $.} according to,
$\Box =-\frac{1}{R^2}\hat\Box=-\frac{1}{R^2} {\mathcal K}^2 $,
whose action on the states is given by \cite{Kim_Noz}
\begin{equation}
{\mathcal K}^2 \vert K, l, m \rangle = K(K+2)
\vert K, l, m \rangle.
\label{Casimir_O4}
\end{equation}
 Here, the $O(4)$ states have been equipped by the quantum numbers, 
$K$, $l$, and $m$ defining the eigenvalues of the respective  
four--, three-- and two--dimensional angular momentum operators
upon same states.
 These quantum numbers 
correspond to the $O(4)/O(3)/O(2)$ reduction chain and
satisfy the branching rules,
$l=0,1,2,.. K$, and  $m=-l,..., +l$. 

Therefore, the corresponding energy spectrum has to be
\begin{equation}
E_K^{(c=0)}(\kappa )=\kappa \frac{\hbar ^2}{2\mu } 
K(K+2).
\label{energy_O4}
\end{equation}
When cast in terms of  $n=K+1$,
the latter spectrum takes the form
\begin{equation}
E_n^{(c=0)}(\kappa )=\kappa \frac{\hbar ^2}{2\mu } 
\left( n^2-1\right),
\label{energy_O(4)}
\end{equation}
which coincides (up to an additive constant)
 with the spectrum of  a particle confined within
an infinitely deep spherical quantum-box well.
Then $n$ acquires meaning of  principal quantum number.

The solutions of eq.~(\ref{chi_free}) are text-book knowledge  
\cite{Kim_Noz} ,\cite{Tjon}
and rely in the following way upon the Gegenbauer polynomials, 
$C_{m}^{\alpha }$, the $O(4)$ orthogonal polynomials,  
\begin{equation}
{\mathcal S}_{Kl}(\chi ,\kappa )=
\sqrt{\kappa}  2^{l+1}l!\sqrt{
\frac{\kappa (K+1)(K-l)!}{2\pi^2 (K+l+1)!}} 
\sin^l\chi C_{K-l}^{l+1}(\cos \chi ).
\label{free_4_sol}
\end{equation}
The complete solutions to eq.~(\ref{chi_free}) and on the unit
hypersphere are the well known hyper-spherical harmonics given by
\begin{equation}
|Klm>=Z_{Klm}(\chi, \theta, \varphi )= {\mathcal S}_{Kl}(\chi ,\kappa=1 )
Y_l^m(\theta ,\varphi),
\label{hypersph_harm}
\end{equation}
where $Y_l^m(\theta ,\varphi)$ are the standard spherical harmonics in 
ordinary three space.

\subsection{Confinement as  $\cot \chi $ barrier  }
Various potentials in conventional flat $E_3$ space
appear as images to the $\cot \chi $  potential in 
eq.~(\ref{harm_4_pot}).
Their explicit forms are determined by  the choice of 
coordinates on $S_R^3$ which shape the line element, $\mbox{d}s$. 
The general expression of the line element in the space under 
consideration and in hyper-spherical coordinates, 
$\Omega =\lbrace\chi, \theta, \varphi \rbrace $, reads
\begin{eqnarray}
\mbox{d}s^2&=&\frac{1}{\kappa}\lbrack \mbox{d}\chi^2 +
\sin^2\chi (\mbox{d}\theta ^2
+\sin^2\theta \mbox{d}\varphi^2)\rbrack.
 \label{metrics_chi}
\end{eqnarray}
Upon the variable substitution, $\chi = f\left( r\right)$, restricted to
$0 \leq f(r)\leq \pi$, 
eq.~(\ref{metrics_chi})
takes the form 
\begin{eqnarray}
\mbox{d}s^2&=&\frac{1}{\kappa } 
\lbrack (f^\prime (r))^2 \mbox{d}r^2 + \sin^2 f (r)
(\mbox{d}\theta^2 +\sin^2\theta \mbox{d}\varphi ^2)\rbrack ,\nonumber\\
&\equiv & D^2(r,\kappa )\frac{\mbox{d}r^2}{r^2} + {\mathcal R}^2 (r,\kappa )
(\mbox{d}\theta^2 +
\sin^2\theta \mbox{d}\varphi ^2),\\
D(r,\kappa )\equiv \frac{ r}{\sqrt{\kappa}} f^\prime (r), &\quad & 
{\mathcal R}(r,\kappa ) \equiv \frac{\sin f(r)}{\sqrt{\kappa}}\,,
\label{metrics_r}
\end{eqnarray}
where $D(r,\kappa )$, and  ${\mathcal R}(r, \kappa )$ 
are usually referred to as 
``gauge metric tensor'' and ``scale factor'', respectively \cite{Ismst}.

Changing  variable in  eq.~(\ref{chi_world}) correspondingly
is standard and  various choices for  $f(r)$  give rise to a variety 
of radial equations in ordinary
flat space with effective potentials which are not even necessarily central.
All these equations, no matter how different that may look,
are of course equivalent, they have same spectra,
and the transition probabilities between the levels  
are independent on the choice for $f(r)$.
Nonetheless, some of the scenarios provided by the different choices for $f(r)$
can be more efficient in the description
of particular phenomena than others.

\begin{quote}
Precisely here lies the power of the curvature concept
as the common prototype of confinement phenomena of different disguises.
In the following we shall present two typical examples for $f(r)$.
\end{quote}

\subsubsection{The $D(r,\kappa )=\frac{r\sqrt{\kappa }}{1+ r^2\kappa }$ 
gauge and a gradient 
dependent confinement potential with a position dependent reduced mass}
A prominent choice 
for the transformation of the angular $\chi $ variable
to the $r$ variable 
has been made in ref.~\cite{qtm_dots}
for the purpose of quantum dots physics. 
This gauge is of general interest in so far as in flat $E_3$ space it describes
a particle with position and curvature dependent mass moving within a 
confinement potential
whose infinite barrier is generated by gradient terms. 
The transformation under consideration reads
\begin{equation}
\chi =\tan^{-1} r\sqrt{\kappa }, \quad 0\leq r\sqrt{\kappa}< \infty\ ,
\label{gauge_dots}
\end{equation}
and corresponds to a parametrization of the ``upper'' hemisphere 
in terms of tangential  projective coordinates with respect to 
the ``North'' pole. 
The line element in this gauge becomes
\begin{equation}
\mbox{d}s^2 =
\frac{1}{(1+r^2\kappa)^2 }r^2 \mbox{d}r
+\frac{1}{\kappa }\sin \left( \tan^{-1}r\sqrt{\kappa }\right)
(\mbox{d}\theta ^2 +\sin\theta \mbox{d}\varphi ^2).
\label{line_dots}
\end{equation}
The intriguing aspect of this gauge is that  in 
the $r$ variable the $\cot \chi$  potential on $S^3_R$ is portrayed
by a gradient dependent potential with
a position and curvature dependent reduced mass.
Indeed, changing the $\chi$ variable in the principal curved space
Schr\"odinger wave equation in (\ref{chi_world}) in accordance 
with eq.~(\ref{gauge_dots})
and upon the substitution \footnote{
This substitution ensures that the $\Phi (r,\kappa )$'s are normalized 
as wave functions  in $ E_3$.
},
\begin{equation}
\psi(\tan^{-1} r\sqrt{\kappa })= (1+\kappa r^2)\Phi (r,\kappa ),
\label{var_chng}
\end{equation}
amounts after some straightforward algebra to
\begin{eqnarray}
-\frac{\hbar^2}{2\mu }\left( 1+\kappa r^2 \right)
{\Big[}(1+\kappa r^2)\frac{\partial^2}{\partial r^2}  
&+&\frac{2}{r}(1+3\kappa r )\frac{\partial }{\partial r}
+6\kappa 
-\frac{l(l+1)}{r^2}{\Big]}
\Phi( r, \kappa )\nonumber\\
+\alpha \cot(\tan^{-1} r\sqrt \kappa )\Phi(r,\kappa ) &=&
E(\kappa )\Phi (r,\kappa ), \quad 
\alpha =\frac{e^2Z}{\epsilon}.
\label{qdts_2}
\end{eqnarray}
Now introducing the position and curvature  dependent mass as
\begin{equation}
\mu^\ast (r,\kappa )=\frac{\mu}{1+\kappa r^2},
\label{mass_dpndt}
\end{equation}
plotted in Fig.~\ref{eff_mss}, and accounting for the relation, 
$\cot \left( \tan^{-1}r\sqrt{\kappa}\right) = 1/(r\sqrt{\kappa})$,
allows to cast eq.~(\ref{qdts_2}) into the following symmetrized  form
of the kinetic terms,
\begin{eqnarray}
{\Big[} \frac{1}{2}
\left(  \frac{1}{\mu^\ast (r,\kappa 
)}\Delta_r +\Delta_r \frac{1}{\mu^\ast (r, \kappa )}\right)
+ v\left(r,\frac{\partial}{\partial r},\kappa \right){\Big]} 
\Phi (r,\kappa )&=&
E(\kappa )\Phi (r,\kappa ).
\label{break}
\end{eqnarray}
The explicit expression for the gradient potential reads:
\begin{eqnarray}
v\left(r,\frac{\partial }{\partial r}, \kappa \right)=
\frac{\alpha }{r\sqrt{\kappa}} -\frac{\hbar^2\kappa }{2\mu^\ast }
\left[  \left( r\frac{\partial }{\partial r}\right)^2
+3 \left( r\frac{\partial }{\partial r}\right)
+3 \right] -\frac{\hbar^2\kappa^2}{\mu }r^2
\left( r\frac{\partial }{\partial r}+1  \right),&&\nonumber\\
\Delta_r= \frac{\partial^2}{\partial  r^2} +\frac{2}{r}
\frac{\partial }{\partial r} -
\frac{l(l+1)}{r^2}\,.&& 
\end{eqnarray}
Equation (\ref{break})  
describes  particles with  
position and curvature dependent masses confined within 
a gradient  potential, a scenario suited for
the case of electrons confined in semi-conductor quantum dots.
\begin{figure}[b]
\center
\includegraphics[width=10.7cm]{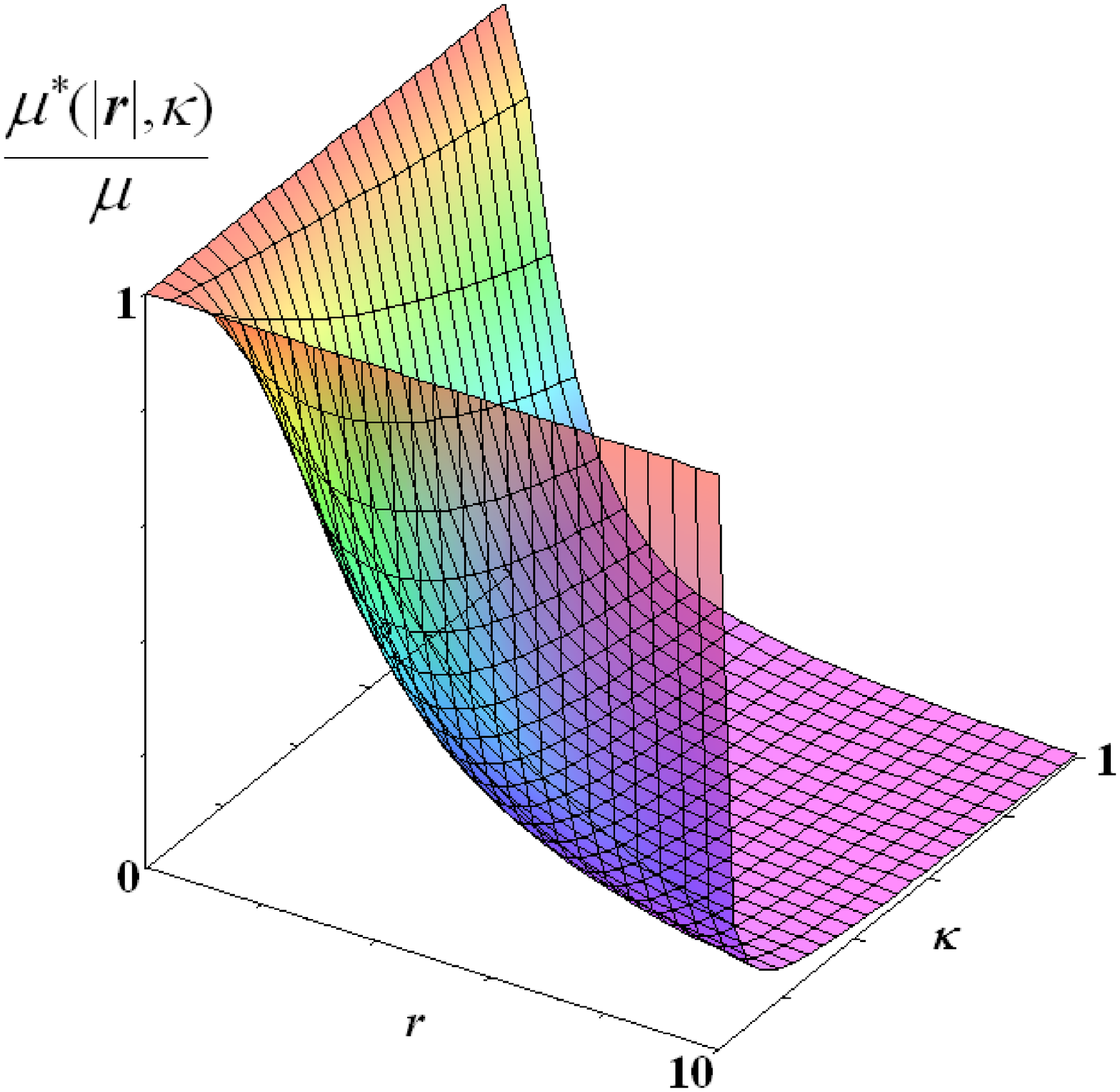}
\caption{ Position and curvature dependence
of the  reduced mass $\mu^*(r,\kappa )$.
Besides effective electron masses in  quantum dots,
one may entertain applicability of this scenario to
evolution of finite valence 
to vanishing parton quark masses as effect of
transversal (relative to the extra dimension)
displacement on the three-dimensional ``plane'' 
tangential to the  ``North'' pole of the hemisphere. 
\label{eff_mss}}
\end{figure}
\begin{quote}
This confinement phenomenon, that occurs due to the 
electron-crystal interaction, may not 
restrict to quantum dots alone. Also 
quarks with position dependent masses due to quark-sea interaction 
may be of interest too \cite{APS}.
\end{quote}

\subsubsection{The $D(r,\kappa )=\pi r  $ 
gauge and the central trigonometric Rosen-Morse 
potential}

An especially simple and convenient parametrization of the $\chi$ variable  
in terms of  $r$, 
also used by Schr\"odinger \cite{Schr40} and corresponding to the $D(r)=\pi r$
gauge  is
\begin{equation}
\chi =\frac{r}{R}\pi \equiv \frac{r}{d}, \quad d=\frac{R}{\pi }, \quad 
\frac{r}{R}\in [0, 1 ], \quad \kappa \to \widetilde{\kappa}= \frac{1}{d^2},
\label{Schr_choice}
\end{equation} 
in which case the line element takes the form
\begin{equation}
\mbox{d}s^2=  \frac{\pi^2}{\widetilde{\kappa}} 
\left(  \mbox{d}\left( r\sqrt{\widetilde{\kappa} }\right)^2 + 
\sin^2\left( r\sqrt{\widetilde{\kappa }}\right)
(\mbox{d}\theta ^2 +\sin^2\theta \mbox{d}\varphi ^2)\right) .
\label{line_pir}
\end{equation}
Here, the length parameter $d$ assumes the r\'ole of 
rescaled hyper-radius.  Correspondingly,
in this particular gauge, the place of the
genuine curvature, $\kappa =1/R^2$, is taken  by  
the rescaled one, $\widetilde{\kappa}=1/d^2$.
Setting now  $c=-2G\sqrt{\widetilde{\kappa} }$,  eq.~(\ref{chi_eq}) 
takes the form of a radial Schr\"odinger equation with
a particular central  potential of infinite depth
(confinement potential)  known in SUSYQM 
under the name of the
{\it  trigonometric Rosen-Morse potential (or, Rosen-Morse I)\/}~\cite{Levai}.
This equation reads
\begin{eqnarray}
{\Big[} 
-\widetilde{\kappa }\frac{\hbar ^2}{2\mu}\frac{\mbox{d}^2 }{\mbox{d}
\left( r\sqrt{\widetilde{\kappa}}\right)  ^2}
&+& {\mathcal V} \left(r\sqrt{\widetilde{\kappa }},
\widetilde{ \kappa} \right){\Big]}  
X\left(r\sqrt{\widetilde{\kappa}},\widetilde{\kappa} \right)
=\left(E(\widetilde{\kappa} ) +\frac{\hbar^2}{2\mu}\widetilde{\kappa} \right) 
X\left(r\sqrt{\widetilde{\kappa}}, \widetilde{\kappa}  \right),\nonumber\\
{\mathcal V} \left(r\sqrt{\widetilde{\kappa} },
\widetilde{\kappa} \right)  &=&
\widetilde{\kappa} \frac{\hbar^2}{2\mu } 
\frac{l(l+1)}{\sin^2 \left( r\sqrt{\widetilde{\kappa} }\right) }
-2G\sqrt{\widetilde{\kappa}}\cot \left( r\sqrt{\widetilde{\kappa} }\right).
\label{ros_morse}
\end{eqnarray}
\begin{quote}
Important to note, SUSYQM suppresses the curvature dependence of
${\mathcal V}$ by absorbing it into the constants and the variable 
through the replacements,
$G\sqrt{\widetilde \kappa }\to b$, and $\widetilde{\kappa}l(l+1)\to a(a+1)$,
and refers to $r\sqrt{\widetilde \kappa }$ as to a dimensionless
position variable, $r\in [0,\pi ]$. 
\end{quote}
Besides Schr\"odinger, eq.~(\ref{ros_morse})
has  been solved by various authors using different schemes.
The solutions obtained in \cite{Stevenson} are built on top 
of Jacobi polynomials of imaginary arguments and parameters that are
complex conjugate to each other,
while ref.~\cite{Vinitski} expands the
wave functions of the interacting case in the free particle basis.
The most recent construction in our previous work  \cite{Quiry} instead
relies upon real Romanovski polynomials. In the $\chi $ variable
and according to eq.~(\ref{trsfm}) (versus 
$\psi (r)=r {\mathcal R}(r)$ 
in $E_3$) 
our solutions take the form, 
\begin{eqnarray}
X_{(K l)}(\chi , \widetilde{\kappa }  ) =  N_{( K l) } 
\sin^{K+1} \chi  e^{-\frac{b\chi }{K+1}  }
R_{K-l }^{(\frac{2b}{K+1 },-( K+1) )}\left(\cot \chi  
\right),
&\quad&  b=\frac{2\mu G}
{\sqrt{\widetilde{\kappa}}\hbar^2}.
\nonumber\\
K=0,1,2,...,\quad  l=0,1,...,K ,&&
\label{Rom_pol}
\end{eqnarray}
where  $N_{(K l)}$ is a normalization constant. 
The 
$R_{n}^{(\alpha, \beta )}(\cot \chi )$ 
functions are the non-classical 
Romanovski polynomials \cite{routh,rom} which are defined by the
following Rodrigues formula,
\begin{eqnarray}
R_{n}^{(\alpha, \beta )}(x)&=&e^{\alpha  \cot^{-1} x} (1+x^2)^{-\beta +1}
\nonumber\\
&&\times \frac{\mbox{d}^n}{\mbox{d}x^n}e^{-\alpha  
\cot^{-1} x} (1+x^2)^{\beta -1 +n},
\label{Rodrigues}
\end{eqnarray}
where $x=\cot r\sqrt{\widetilde{\kappa}} $ 
(see ref.~\cite{raposo} for a recent review).

The energy spectrum of ${\mathcal V}\left(r\sqrt{\widetilde{\kappa} },
\widetilde{\kappa } \right)$ 
is given by
\begin{equation}
E_{K}(\widetilde{\kappa} )=-\frac{G^2}{\frac{\hbar^2}{2\mu}}   
\frac{1}{(K+1)^2}
+ \widetilde{\kappa} \frac{\hbar ^2}{2\mu }( (K+1)^2-1), \quad l=0,1,2,...,K.
\label{enrg_cot}
\end{equation}
Giving $(K+1)$ the interpretation of a principal quantum number
 $n=0,1,2,...$ (as in the $H$ atom),
one easily recognizes that the energy in eq.~(\ref{enrg_cot}) is
defined by the Balmer term and its inverse of opposite sign, thus
revealing $O(4)$ as dynamical symmetry of the problem.
Stated differently, particular  levels 
bound within different  potentials (distinct by the values of $l$) 
carry same energies and align to levels 
(multiplets) characterized, similarly to the free case in 
eq.~(\ref{Casimir_O4}), by the four dimensional angular momentum,
$K$. The  $K$-levels belong to the irreducible  
$O(4)$ representations of the type  $\left(\frac{K}{2},\frac{K}{2} \right)$. 
When the confined particle carries spin-1/2, as is
the case of electrons in quantum dots, or quarks in baryons, 
one has to couple the spin, i.e. the
$\left(\frac{1}{2},0 \right)\oplus \left(0,\frac{1}{2} \right)$ representation,
 to the previous multiplet,
ending up with the (reducible) $O(4)$ representation
\begin{equation}
\vert K,l,m, s =\frac{1}{2}\rangle = \left(\frac{K}{2},\frac{K}{2} \right)
\otimes \left[
   \left(\frac{1}{2},0 \right)\oplus \left(0,\frac{1}{2} \right)\right].
\label{cluster-K}
\end{equation}
This representation contains  $K$ parity dyads
and a state
of maximal spin, $J_{\mathrm{max}}=K+\frac{1}{2}$, without
parity companion and  
of either positive ($\pi =+$) or, negative ($\pi =-$) parity,
\begin{equation}
\frac{1}{2}^\pm , ...,\left( K-\frac{1}{2}\right)^\pm, 
\left( K+\frac{1}{2}\right)^\pi \in \vert K,l,m, s=\frac{1}{2}\rangle .
\label{filon}
\end{equation}
As we shall see below this scenario turns to be the one adequate for
the description of non-strange baryon structure.\\

\noindent
\begin{quote}
The above examples are illustrative of the power of curved space potentials 
as sources for a variety of effective confinement
potentials in ordinary flat $E_3$ space.
\end{quote}
However, much care is in demand when working with such potentials.
One should be aware of the fact that physical observables do not
depend on the gauge chosen and performing in the $r$ variable has to
be consistent with performing in the $\chi $ variable. 
We shall come back to this point in  the next section.

\section{ $S_R^3$ potentiology:The baryons}

\subsection{The nucleon spectrum in the $SO(4)$ symmetry scheme }
The spectrum of the nucleon continues being enigmatic despite 
the long history of the respective studies (see refs.~\cite{Lee}, 
\cite{Afonin} for recent reviews).
Unprejudiced inspection of the data reported by the Particle
Data Group \cite{PART} reveals a systematic degeneracy  of the
excited states of the baryons of the best coverage, the
nucleon $(N)$ and the $\Delta (1232)$. 
Our case is that
\begin{itemize}
\item  levels and level splittings of the nucleon and $\Delta $ spectra
match the spectrum of in eq.~(\ref{enrg_cot}),

\item the curvature induced confinement potential in eq.~(\ref{ros_morse})
 is the exactly solvable extension of
to the Cornell potential predicted by Lattice QCD.

\end{itemize}

\begin{enumerate}
\item \underline{The $N$ and $\Delta$ spectra:}
Afreque
ll the observed nucleon resonances with masses below $2.5$ GeV
fall into the three $K= 1,3,5$ levels in eq.~(\ref{filon})
with only the two 
$F_{17}$ and $H_{1,11}$ states still ``missing'',
an observation due to refs.~\cite{MK-97}.
Moreover, the level splittings follow with an amazing accuracy 
eq.~(\ref{enrg_cot}). This is in fact a result already reported in
our previous work in ref.~\cite{Quiry_2} where we
assumed dominance of a quark-diquark configuration in nucleon structure,
and fitted the nucleon spectrum in Fig.~\ref{levels-nucleon}
to the spectrum of the trigonometric
Rosen-Morse potential (c.f. eq.~(\ref{ros_morse})) 
by the following set of parameters,
\begin{eqnarray}
&\mu= 1.06 \,\, \mathrm{fm}^{-1}\ ,\quad G=237.55\  
\mbox{MeV}\cdot\mbox{fm} \ ,\quad d=2.31 
\ \mathrm{fm}.
\label{parameters}
\end{eqnarray}

However, in ref.~\cite{Quiry_2} the curvature concept has not been 
taken into consideration and because of that  the $d$ quantity 
did not have any deeper meaning but the one of
some length matching parameter. This contrasts  the present work, which
in being entirely focused on the geometric aspect of confinement,
places $d$ on the firmer ground of a parameter encoding a space curvature.

\begin{figure}[b]
\center
\includegraphics[width=10.5 cm
]{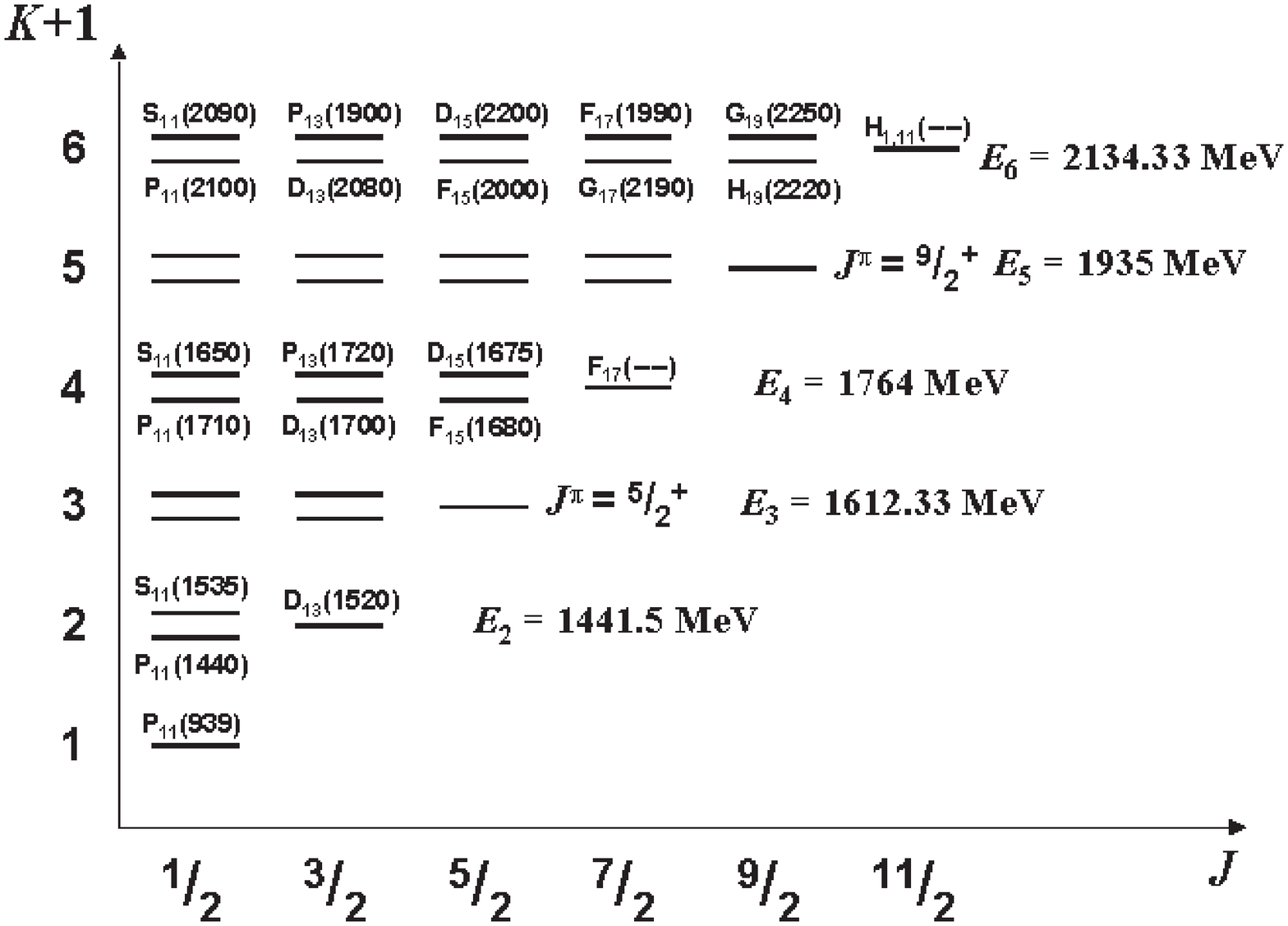}
\caption{Assignments of the reported $N$ excitations to
the $K$ levels of the $S_R^3$ potential,
 ${\mathcal V}(r\sqrt{\widetilde{\kappa}},\widetilde{ \kappa} )$,
in eq.~(\ref{ros_morse}), taken as the quark-diquark confinement potential.
The potential parameters are those from eq.~(\ref{parameters}).
Double bars represent parity dyads, single bars the unpaired states of maximum
spin. The notion $L_{2I,2J}(--)$ has been used  for resonances  ``missing'' 
from a level. 
The model predicts two more levels of maximal spins $J^\pi =5/2^+$, 
and  $J^\pi =9/2^+$, respectively, which are completely  ``missing''. 
In order not to overload the figure with notations, the names of
the resonances belonging to them have been suppressed.
The predicted energy at rest (equal to the  mass) 
of each level is given to its  most right.  
\label{levels-nucleon}}
\end{figure}

\noindent
Almost same set of parameters, up to a modification of $d$ to  
$d= 3\, \mbox{ fm}$, fits the $\Delta (1232)$ spectrum, 
which exhibits exactly same  degeneracy patterns, and
 from which only the three $P_{31}, P_{33}$, and $D_{33}$ states 
from the $K=5$ level are ``missing''. Remarkably, none of the reported states, 
with exception of the $\Delta (1600)$ resonance, 
presumably a hybrid, drops from the systematics.
The unnatural parity of the $K=3,5$ levels requires a pseudoscalar
diquark. For that one has to account for
an $1^-$ internal excitation of the diquark which, when coupled to
its maximal spin $1^+$, can produce a pseudoscalar 
in one of the possibilities. The change of parity from natural to
unnatural can be given the interpretation of a  chiral phase
transition in baryon spectra.   
Levels with $K=2,4$  have been  attributed to 
entirely  ``missing'' resonances in both the $N$ and $\Delta $ spectra.
To them, natural parities have been assigned on the basis of a
detailed analysis of the $1p-1h$ Hilbert space of three quarks and
its decomposition in the $\vert K, l, m, s=\frac{1}{2} \rangle $ 
basis \cite{MK-97}.
The $\Delta (1232)$ spectrum obtained in this way is shown in 
Fig.~\ref{levels-Delta}.
We predict a total of 33 unobserved resonances of a dominant
quark-diquark configurations  in the  $N$ and $\Delta (1232)$
spectra with masses below $\sim 2500$ MeV, much less but any other
of the traditional models.

In ref.~\cite{Quiry_2} mentioned above, the 
potential in eq.~(\ref{ros_morse}) has been considered 
in the spirit of SUSYQM as a 
{\it central two-parameter \/} potential in $E_3$ and 
without reference to $S_R^3$, a reason for which the 
values of the parameter accompanying the
$\csc^2$ term had to be taken as integer 
{\it ad hoc\/} and for the only sake of a better fit to
the spectra, i.e., without any deeper justification.
Instead, in the present work, 
\begin{quote}
we fully recognize
that  the higher dimensional potential
${\mathcal V}(\chi, \kappa )$ in eq.~(\ref{chi_world}),
which acts as the prototype of Rosen-Morse I,
is a {\it non-central one-parameter\/} potential 
in which the strength of the
$\csc^2$ term, the centrifugal barrier on $S_R^3$,  is 
uniquely fixed by the eigenvalues of underlying three-dimensional
angular momentum.
\end{quote}

\item\underline{Relationship to QCD dynamics:}
 The convenience of the scenario under consideration
is additionally backed by the fact that the Cornell 
quark confinement potential \cite{Cornell} predicted by Lattice QCD
\cite{Lattice} is no more but the small-angle approximation to
$\cot r\sqrt{\widetilde{\kappa} }$ in eq.~(\ref{harm_4_pot}).
Indeed, the first terms of the series expansion are
\begin{eqnarray}
-2G\sqrt{\widetilde{\kappa}}\cot r\sqrt{\widetilde{\kappa} }
+ \widetilde{\kappa} \frac{\hbar^2}{2\mu } 
\frac{l(l+1)}{\sin^2 \left( r\sqrt{\widetilde{\kappa} }\right) }
&\approx&
-\frac{2G}{r} +\frac{2G\widetilde{\kappa}}{3} r
+ \frac{\hbar^2}{2\mu } 
\frac{l(l+1)}{r^2 },
\label{crnl}
\end{eqnarray}
with $\widetilde{\kappa}=\frac{1}{d^2}=\frac{\pi^2}{R^2}$.
\begin{quote}
Therefore, ${\mathcal V}(r\sqrt{\widetilde{\kappa}},\widetilde{ \kappa} )$,
is the  exactly solvable 
{\it \underline{t}rigonometric  \underline{e}xtension
to the {\underline{C}ornell}\/} potential, 
a reason for which we shall frequently refer
to ${\mathcal V}(r\sqrt{\widetilde{\kappa}},\widetilde{ \kappa} )$ as
TEC potential.
\end{quote}

\end{enumerate}

Finally, the ``curved'' $1/r$ potential has been completely 
independently used in \cite{Ismst} within the context of charmonium physics.
\begin{figure}[b]
\center
\includegraphics[width=10.5 cm
]{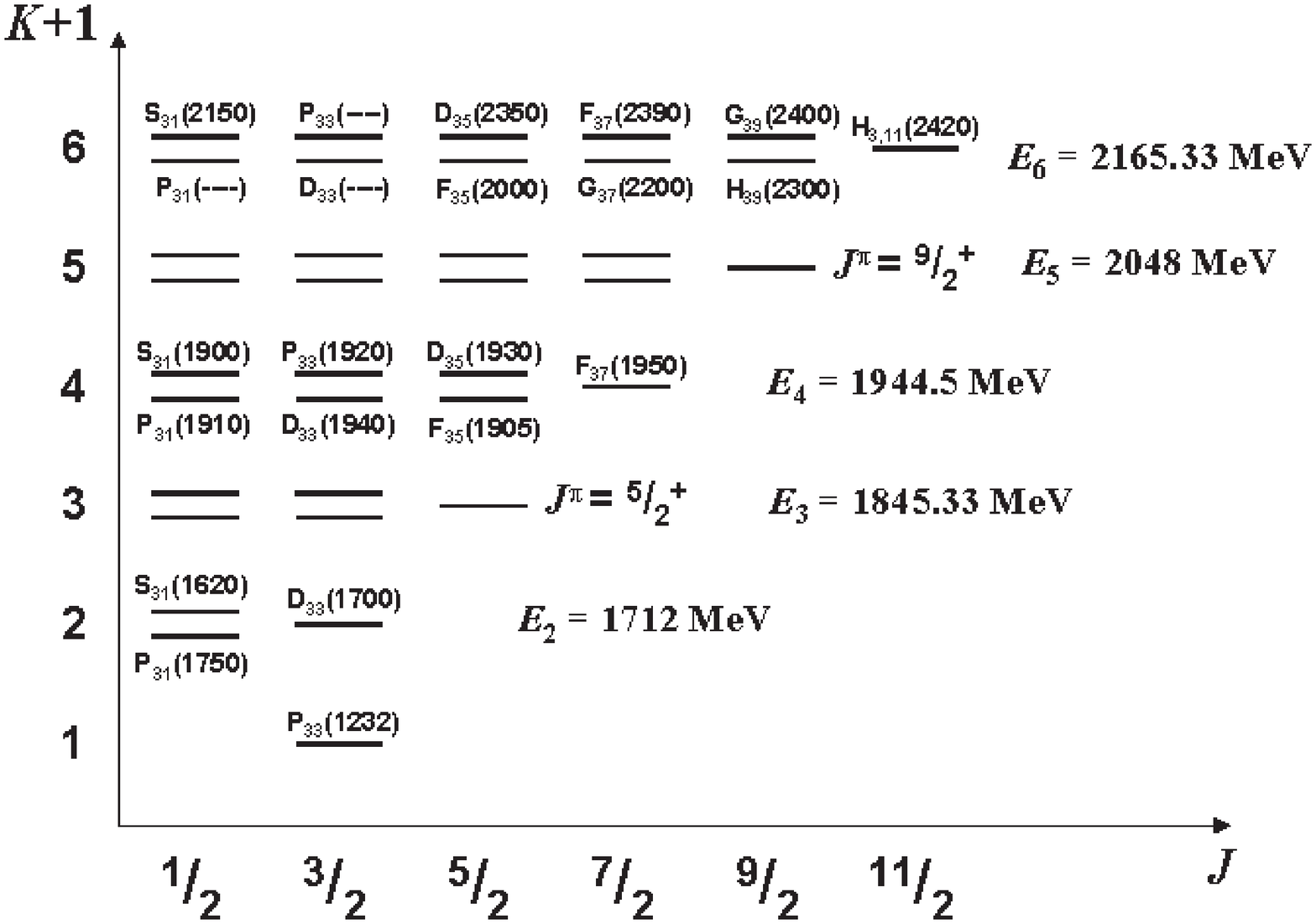}
\caption{Assignment of the reported $\Delta (1232)$ excitations to
the $K$ levels of the $S_R^3$ potential ${\mathcal V}
(r\sqrt{\widetilde{\kappa}},\widetilde{ \kappa} )$
in eq.~(\ref{ros_morse}), taken as the quark-diquark confinement potential.
The potential parameters are those from eq.~(\ref{parameters})
with exception of the $d$ value taken here as $d=3$ fm.
Other notations same as in Fig.~\ref{levels-nucleon}.
\label{levels-Delta}}
\end{figure}
However, in refs.~\cite{Ismst} at the end
the degeneracy of the states has been
removed through an {\it ad hoc\/} extension of the Hamiltonian to include
an additional  $L^2/r^2$-term
of sign opposite to four-dimensional centrifugal barrier,
$L^2/\sin^2\chi$, so that
all states with  $l=1,2,...,K$ 
could be pushed below the $S$ state belonging to a given $K$.  
In this way a better description of the charmonium states has been achieved
indeed but on the cost of compromising  consistency of  the
geometric $S_R^3$ concept. 
In contrast to the charmonium,
in $N$ and $\Delta (1232)$  baryon spectra
the $S_R^3$ degeneracy patterns are pretty well pronounced especially
in the most reliable region below 2000 MeV where 
our predicted $K=1,3$ levels in the $\Delta $ spectrum
appear even complete. This occasion
allows to maintain the $S_R^3$ geometric concept intact and the exact
wave functions unaltered, so far.   
In subsection C below these wave functions   
will be put at work in the description of the mean square proton charge radius.

\subsection{The nucleon spectrum in the $SO(2,1)$ symmetry scheme }
The energy spectrum in 
eq.~(\ref{enrg_cot}) can equivalently be cast in terms of the eigenvalues
 of the $SO(2,1)$ Casimir, the pseudo-angular momentum operator,
${\mathcal J}^2=-J_\pm J_\mp +J_0^2\pm J_0$ \cite{Wybourne},\cite{Marcelo}. 
Here, the operators $J_\pm$ 
and $J_0$ satisfy the group algebra
$\lbrack J_+,J_- \rbrack =-2J_0 $, and $\lbrack J_0,J_\pm \rbrack=\pm J_\pm $. 
The eigenstates are labeled by the quantum numbers $j$, and $m^\prime$
which define the respective eigenvalues of ${\mathcal J}^2$ and $J_0$ 
according to $J_0 |j,m^\prime \rangle =m^\prime | j,m^\prime\rangle$, 
and ${\mathcal J}^2 |j,m^\prime \rangle =
 \left(\left(
j- \frac{1}{2}\right)^2 -\frac{1}{4}\right)|j,m^\prime \rangle $, 
respectively. 

 The group $SO(2,1)$, in being non-compact, allows only for infinite 
dimensional unitary  representations labeled as 
$D_{j}^{\pm \left( {m^\prime}\right)}$ where 
the positive, and negative upper signs refer in their turn to $m^\prime $ 
values limited from either below, $m^\prime =j +n$ with $n$ non-negative
integer, or above,
$m^\prime = j-n$ where we used a nomenclature 
of positive integer or half-integer $j$ (also known as Bargmann index). 

Back to eq.~(\ref{enrg_cot}), it is not obvious how to re-express 
the general two-term energy formula containing 
both the quadratic and inverse quadratic eigenvalues of the $SO(4)$ Casimir
in terms of $SO(2,1)$ quantum numbers. The most obvious option consists in
nullifying the potential strength, i.e. setting $G=0$, 
which takes one back to the free particle on
the hypersphere. In this case only the quadratic terms survives which is
easily equivalently rewritten to
\begin{equation}
E_j(\widetilde{\kappa})=\widetilde{\kappa }
\frac{\hbar^2}{2\mu }\left( \left(m^\prime \right)^2 -1\right),
\quad j= l+1, \quad m^\prime =j+n,
\label{G=0}
\end{equation}
where $l$ is the ordinary angular momentum label, while $n$ is the radial
quantum number (it equals the order of the polynomial shaping the wave
function labeled by $K$ in eq.~(\ref{Rom_pol}).
The $m^\prime $ label  is limited from below and the whole spectrum 
can be associated with the basis of the infinite unitary 
$SO(2,1)$ representation, $D^{+ (m^\prime)}_j$.
It is obvious that the degeneracy patterns in the
$SO(2,1)$ spectrum designed in this manner 
are same as the $SO(4)$ ones.

Perhaps nothing expresses the $SO(2,1)/SO(4)$ symmetry correspondence
better but this extreme case in which the manifestly $SO(4)$ symmetric
centrifugal energy on the $(3D)$ hypersphere is cast in terms of
$SO(2,1)$ pseudo-angular momentum values.

Although the bare $l(l+1)\csc^2$ potential is
algebraically in line with   $AdS_5/CFT$ correspondence, it 
completely misses the perturbative aspect of QCD dynamics.
The  better option for getting rid of the inverse-quadratic term in 
eq.~(\ref{enrg_cot}) is to permit  $K$ dependence of 
the potential strength and choose  $G=g(K+1)$ with $g$ being
a new free parameter.
Such a choice (up to notational differences) has been made in \cite{Koca}.
If so, then the energy takes the form 
\begin{equation}
E_{j}(\widetilde{\kappa} )=- g^2{\frac{\hbar^2}{2\mu}}   
+ \widetilde{\kappa} \frac{\hbar ^2}{2\mu }
\left( \left( m^\prime \right)^2-1\right), 
\quad j=  1, 2, 3, ....
\label{enrg_su11}
\end{equation}
The above manipulation does not affect the degeneracy patterns as it only
provokes a shift in the spectrum by a constant.
Compared to eq.~(\ref{G=0}) the new choice allows the former inverse quadratic
term to still keep presence as a contribution to the energy
depending on a free constant parameter, $g$. 
In this manner, the $SO(2,1)$ energy 
spectrum continues being described by a two-term formula, a 
circumstance that allows for a best fit to the $SO(4)$ description.

Once having ensured that the $SO(2,1)$ and 
$SO(4)$ spectra share same degeneracy patterns, 
one is only left with the task to check consistency of the
level splittings predicted by the two schemes.
Comparison of eqs.~(\ref{enrg_cot}) and (\ref{enrg_su11}) shows that
for the high-lying levels where the inverse quadratic term becomes negligible,
both formulas can be made to coincide to high accuracy by a proper choice 
for $g$.  That very  $g$ parameter can be used once again to
fit the low lying levels to the $SO(4)$ description, 
now by  a value possibly different from the previous one.

This strategy allows to make the $SO(2,1)$ and $SO(4)$ descriptions of 
non-strange baryon spectra sufficiently close and establish 
the symmetry correspondence. In that manner
we  confirm our statement quoted in the introduction that the TEC 
potential is in line  with  both the algebraic aspects of 
$AdS_5/CFT$ and QCD dynamics and provides a bridge between them.

\subsection{The proton mean square charge radius}
In this section we shall test the potential parameters
in eq.~(\ref{parameters}) and the wave function in 
eqs.~(\ref{trsfm}),~(\ref{Rom_pol})
in the calculation of the proton electric form-factor, the touch stone 
of any spectroscopic model. As everywhere through the paper, the
internal nucleon structure is  approximated by a
quark-diquark configuration.
In conventional three-dimensional flat space the electric 
form factor is defined 
in the standard way \cite{Roberts_rev} as
the  matrix element of the charge component,
$ J_0(\mathbf{r})$, of the proton electric current 
between the states of the incoming, $\mathbf{p}_i$, and 
outgoing, $\mathbf{p}_f$, electrons in the dispersion process,  
\begin{equation}
G_{\mathrm{E}}^{\mathrm{p}}( |\mathbf{q}|)=
<\mathbf{p}_f| J_0(\mathbf{r})|\mathbf{p}_i>,
\quad \mathbf{q}=\mathbf{p}_i-\mathbf{p}_f.
\label{el_ff}
\end{equation}
The mean square charge radius is then defined in terms of the slope
of the electric charge form factor at origin and reads,
\begin{equation}
\langle \mathbf{r}^2\rangle =
-6\frac{\partial
G_{\mathrm{E}}^{\mathrm{p}}( |\mathbf{q}| ) }{\partial |\mathbf{q}|^2}{\Big|}_
{|\mathbf{q}|^2=0}\,.
\label{chrg_rd}
\end{equation}
On $S_R^3$, the three-dimensional radius vector, $\mathbf{r}$, has to
be replaced by, $\mathbf{\bar r}$ with $|\mathbf{\bar r}|=R\sin\chi
=\sin\chi/\sqrt{\kappa } $ 
in accordance with eqs.~(\ref{4d_sphr}).
The  evaluation of eq.~(\ref{el_ff}) as four-dimensional
Fourier transform requires  the four-dimensional plane wave,
\begin{equation}
e^{i q\cdot \bar x}= 
e^{i|\mathbf{q}||\mathbf{\bar r}|\cos \theta }=
e^{i{|\mathbf{q}|}\frac{\sin \chi}{\sqrt{{\kappa}}} 
\cos \theta },
\quad |\mathbf{\bar r}|=R\sin\chi=\frac{\sin \chi}{\sqrt{{\kappa}}}.
\label{4_PW_FF}
\end{equation}
The latter refers to a $z$ axis chosen along the
momentum vector (a choice justified in elastic scattering
\footnote{A consistent definition of the four-dimensional plane wave in
$E_4$ would require an Euclidean $q$ vector. However, for 
elastic scattering processes, of zero energy transfer, where $q_0=0$,
the $q$ vector can be chosen to lie entirely in $E_3$, 
and be identified  with the physical space-like momentum transfer.}), 
and a position vector of the confined quark having in general 
a non-zero  projection on the extra dimension axis in 
$E_4$.  

 The integration volume on $S_R^3$ is given by 
$\sin^2\chi \sin\theta \mbox{d}\chi \mbox{d}\theta \mbox{d}\varphi$.
The explicit form of the nucleon ground state wave function obtained from
eq.~(\ref{Rom_pol}) in the  $\chi$ variable reads
\begin{eqnarray}
X_{(00)}(\chi ,\widetilde{\kappa } )&=&N_{(00)}e^{-b\chi }
\sin \chi ,\nonumber\\
N_{(00)}&=&\frac{4b(b^2+1)}{1-e^{-2\pi b}} \, , \quad b=\frac{2\mu G}
{\sqrt{\widetilde{\kappa}}\hbar^2}.
\label{wafu_gst}
\end{eqnarray}
With that, the charge-density takes the form,
 $J_0(\chi ,\widetilde{\kappa} )=e_p|
\psi_{\mathrm{gst}}(\chi ,\widetilde{\kappa })|^2$,
$e_p=1$. In effect, eq.~(\ref{el_ff})
 amounts to the calculation of the following integral,
\begin{eqnarray}
G_{\mathrm{E}}^{\mathrm{p}}(|\mathbf{q}|, \widetilde{\kappa } )
&=&\sqrt{\kappa} \int_0^{\pi } \mathrm{d}\chi  
\frac{
\left( X_{(00)}(\chi,\widetilde{ \kappa} )\right)^2 
\sin (|\mathbf{q}|\frac{\sin\chi}{\sqrt{\kappa }})}
{|\mathbf{q}|\sin\chi }   ,
\label{el_ff_gst}
\end{eqnarray}
where the dependence of the form factor 
on the curvature has been indicated explicitly.

\begin{figure}[b]
\center
\includegraphics[width=8.5 cm
]{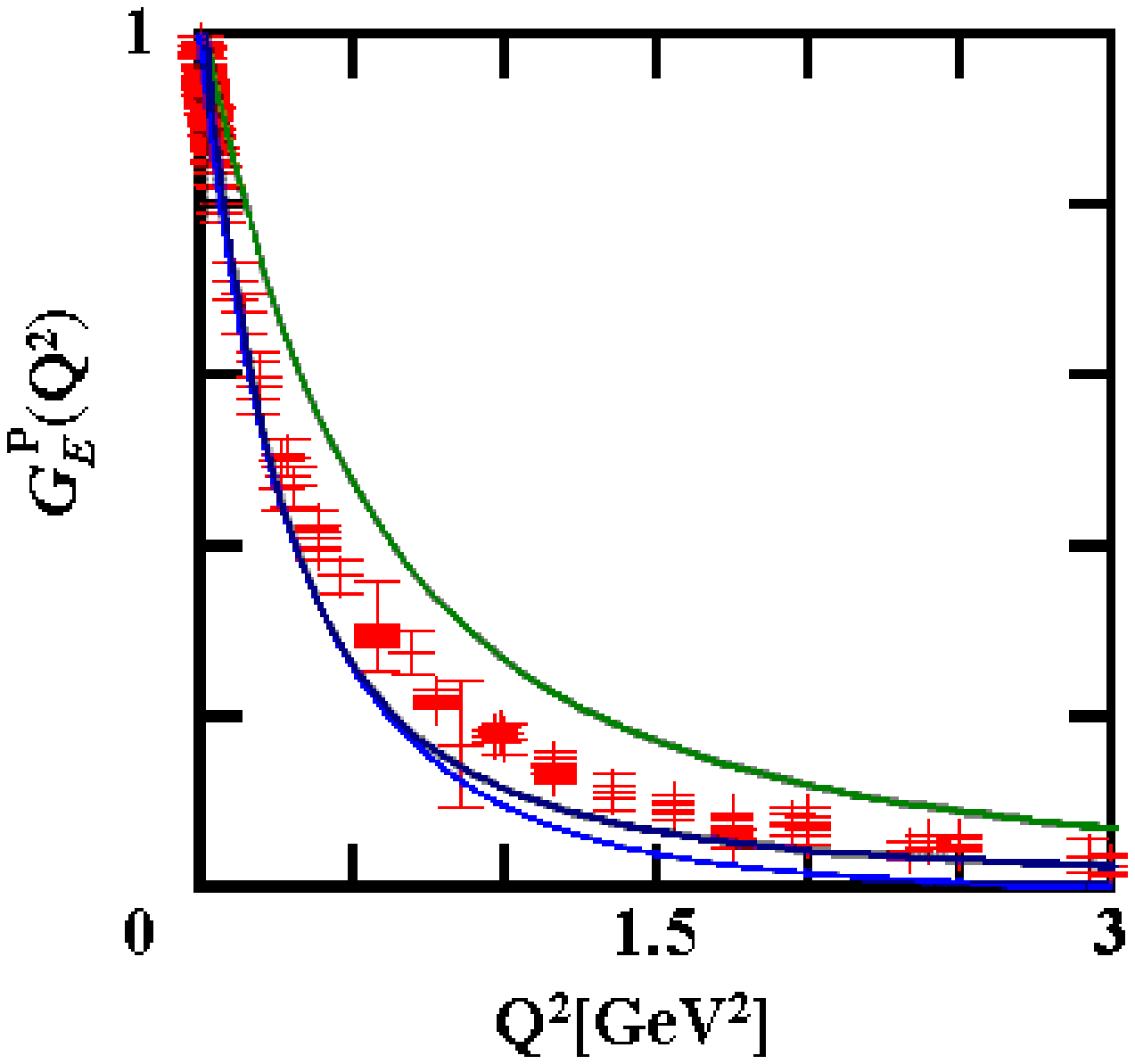}
\caption{The electric charge form factor of the proton calculated for
various curvature parameters. The upper curve corresponds to the curvature
as fitted to the nucleon spectrum, the curvature leading to the
middle curve has been fitted to the experimental value of the mean square 
of the charge radius (see Table 1).
The lowest curve follows from a 
Bethe-Salpeter calculation based upon an instanton induced two-body 
potential and has been presented in ref.~\cite{Metsch}. 
Data compilation taken from \cite{Metsch}. 
\label{charge_FF}}
\end{figure}
The integral is taken numerically and the resulting
charge form factor of the proton is displayed in Fig.~4 together with data.
The best fit values for $R$ (equivalently, $d$), and the related 
curvatures are given in Table \ref{best_fit},
for illustrative purpose.
\begin{table}
\vspace*{0.21truein}
\caption{Best fit values for the $S_R^3$ curvature. }
\vspace{0.21truein}
\begin{tabular}{lcccc}
\hline
~\\
 & hyper-radius $R$  &  $\kappa =\frac{1}{R^2}$&
rescaled  hyper-radius $d=\frac{R}{\pi }$
& $\widetilde{\kappa}=\frac{1}{d^2}$  \\  
~\\  
\hline
\hline
~\\
$N$ spectrum &7.26 fm & 0.019 fm$^{-2}$ & 2.31 fm& 0.187 fm$^{-2}$\\
\hline
~\\
$\Delta $ spectrum & 9.42 fm & 0.011 fm$^{-2}$  & 3 fm& 0.11 fm$^{-2}$ \\
\hline
~\\
form factor  & 10.46 fm& 0.009 fm$^{-2}$ &
3.33 fm &  0.090 fm$^{-2}$\\
\hline
~\\
\end{tabular}
\label{best_fit}
\end{table}
The best fit value of the
mean square charge radius  is found as 
\begin{equation}
\langle \mathbf{\bar r}^2\rangle =0.87 \, \mbox{fm}^2,
\end{equation}
and reproduces well the corresponding experimental value of 
$\langle \mathbf{r}^2\rangle_{\mbox{exp}}= 0.8750$ fm$^2$ \cite{PART}.
We further observe that our best fit is of the quality of the calculation 
of same observable within the framework of the Bethe-Salpeter equation 
based upon an instanton induced two-body potential \cite{Metsch}.\\

\noindent
A comment is in order on how the present resut compares to
our previous work \cite{Quiry_2} where same observable has 
been calculated without reference to the curvature concept.
There, the corresponding  Schr\"odinger equation 
(with essentially same potential) has been written in
terms of the three-dimensional Laplacian versus four-dimensional in the
present work. Using eq.~(\ref{ros_morse}) it can be shown that
the $E_3$ form factor in ref.~\cite{Quiry_2} represents 
the small $\chi $ limit, $\sin\chi \approx \chi $ with 
$\chi=r\sqrt{\widetilde{\kappa}}$, of the  $S_R^3$  case, 
an occasion that allowed to take it in closed form.
Comparing the $E_3$ to the $S_R^3$ calculation reveals the insignificant
differences shown in Figs.~\ref{lvls-Delta}, and ~\ref{lvls-Delta}. 
The coincidence is due to the rapid exponential 
fall of the ground state wave function in eq.~(\ref{wafu_gst})
which strongly damps the large $\chi $ angle contributions to 
the integral in eq.~(\ref{el_ff_gst}) (c.f. Fig.~\ref{Cheese}).     

The result shows that specifically in the $D(r,\kappa )=\pi r$ gauge,
\begin{itemize}
\item performing in $E_3$ is consistent with 
performing on $S_R^3$ in the small $\chi$ angle limit, in which
$\sin \chi \approx \chi$ with $\chi =r\sqrt{\widetilde{\kappa }}$,
\item the proton charge electric 
form-factor is not sufficiently a sensitive observable toward the
curvature parameter.
\end{itemize}

\begin{figure}[b]
\center
\includegraphics[width=9.5 cm
]{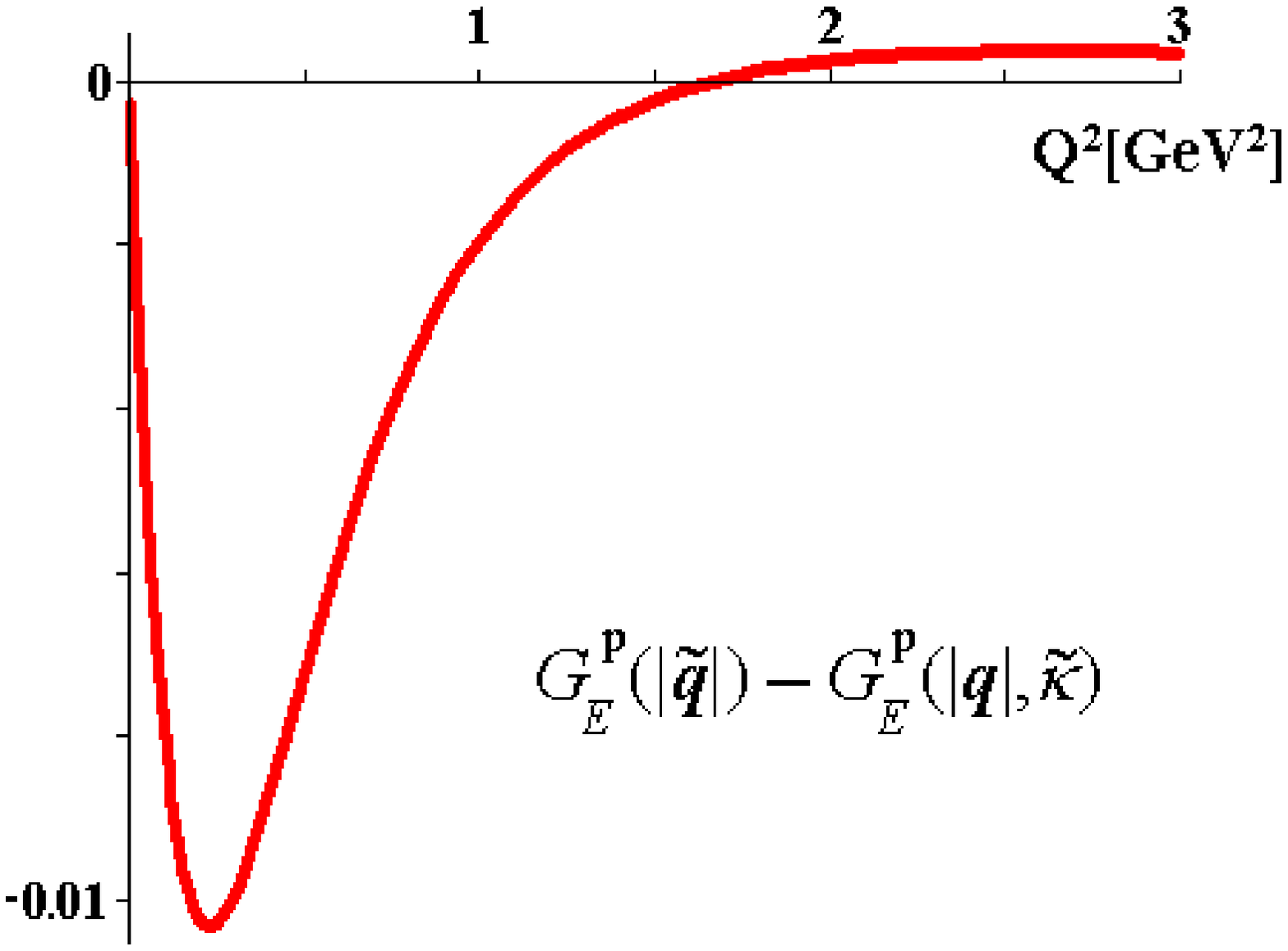}
\caption{
Difference between the electric charge form factor, 
$G_E^p(|\mathbf{q}|, \widetilde{\kappa} )$, calculated in 
$E_4$, with  $\widetilde{\kappa} =1/d^2$, $d=2.31$ fm,
and $G_E^p(|\mathbf{\widetilde q}|)$ with 
$|\mathbf{\widetilde q}|\equiv |\mathbf{q}|d$
calculated in \cite{Quiry_2} in ordinary three space in closed form
for the same $d$ value. 
On the figure the form factors have been plotted as functions of 
$Q^2=-\mathbf{q}^2=|\mathbf{ q}|^2$.
The insignificance of this difference 
illustrates  consistency of the three-dimensional Fourier-transform 
with the small $\chi $ angle 
approximation to the four-dimensional Fourier transform 
in the $D(r,\kappa )=r\pi $ gauge. 
\label{lvls-Delta}}
\end{figure}
This contrasts excited states whose wave functions for $l>0$ 
show non-negligible large $\chi $ angle dependences
(c.f. Fig.~\ref{Cheese}) in which case the
Fourier transforms on $S_R^3$ will become distinguishable from
those in $E_3$.
This is visualized in Fig.~\ref{primouno} by  the electric charge form-factor
for an $l=2$ state from the second observed level with $K=3$ which
corresponds to the first $F_{15}$ resonance. 

\begin{figure}[b]
\center
\includegraphics[width=97 mm, height=69mm
]{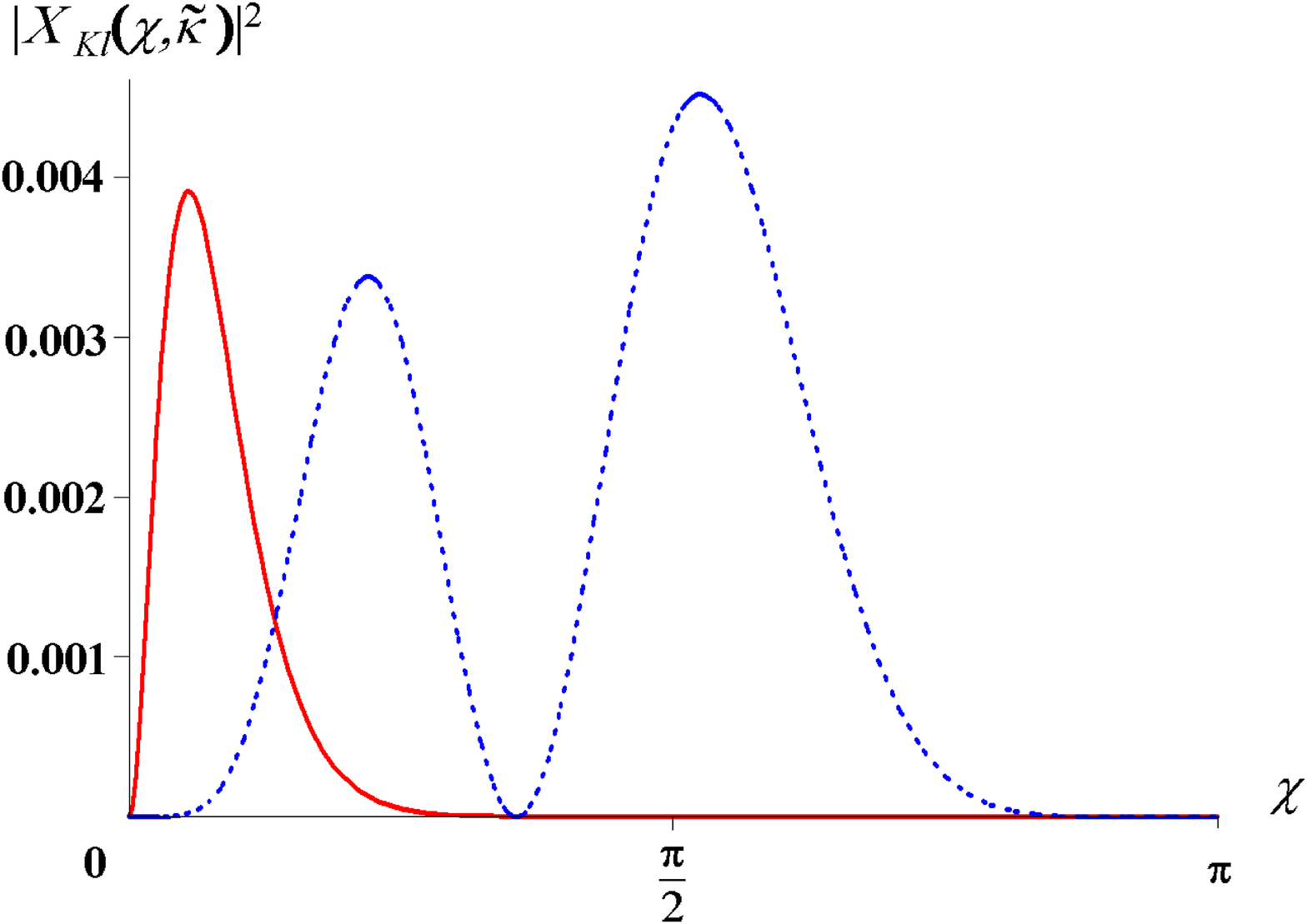}
\caption{ Charge density profiles in the ground state (solid line),
and the $K=3,l=2$ excitation (dashed line),
corresponding to the first $F_{1, 5}$ resonance.
The wave function have been taken unnormalized.
It is visible that while the former damps the small $\chi$ angle contributions
from the four dimensional plane wave, 
the latter captures a significant amount of them which
enables it to distinguish flat from curved spaces.  
\label{Cheese}}
\end{figure}

\begin{figure}[b]
\center
\includegraphics[width=90 mm,height=80mm
]{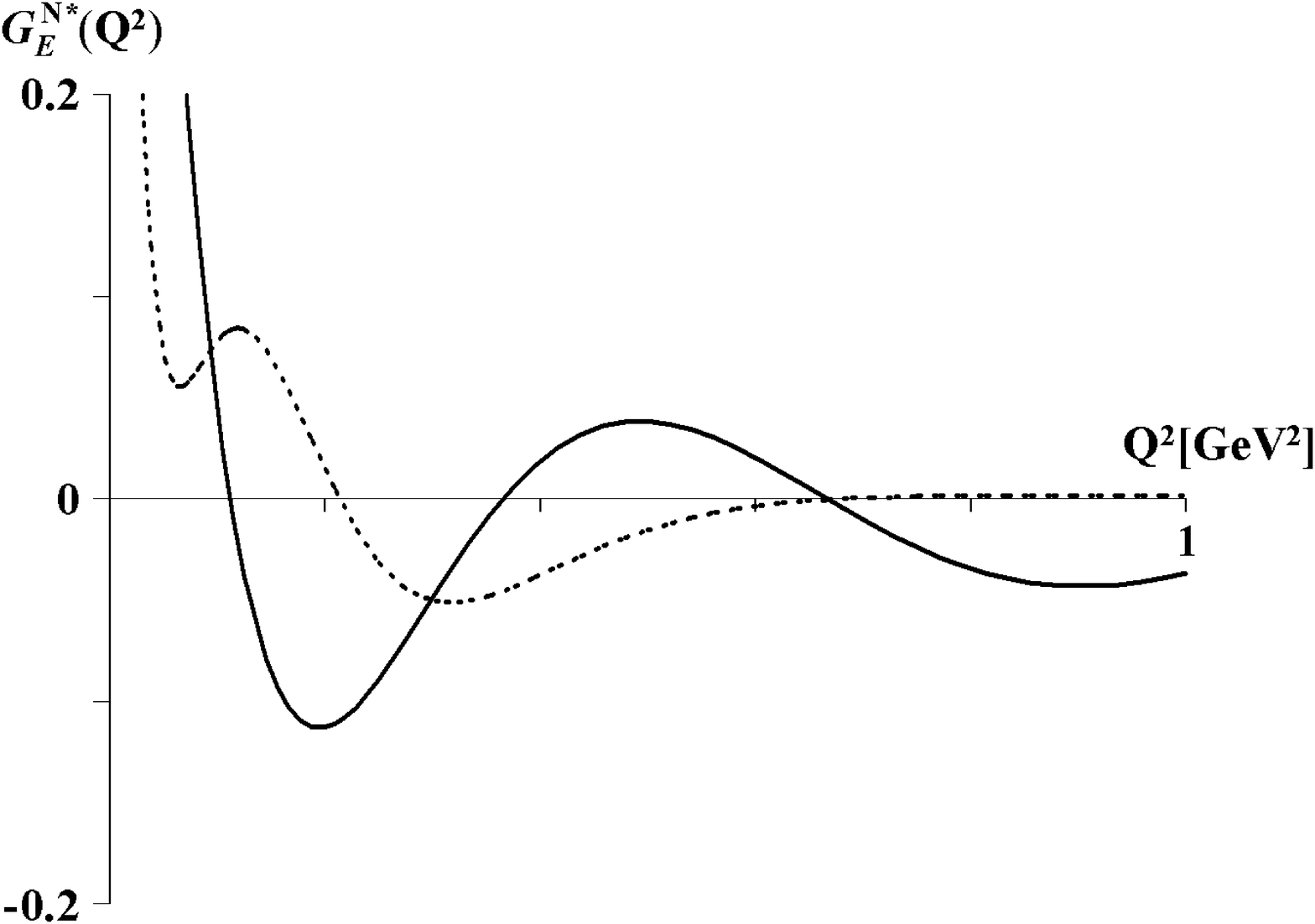}
\caption{ Electric charge form factor of the $K=3$, $l=2$ resonance 
 from plane space (dashed line) versus the $S_R^3$ one (solid line).
 It is visible that in the vicinity of
 $Q^2=1$ GeV$^2$, the curved space predicts
a non-zero form factor versus a vanishing one 
following from flat space. 
\label{primouno}}
\end{figure}

\section{Curvature shut-down:The deconfinement}

The presence of the curvature parameter in the trigonometrically extended
Cornell confinement potential
opens an intriguing venue toward deconfinement as a 
$S_R^3$ curvature shut-down. It can be shown that  
\begin{quote}
high-lying  bound 
states from the trigonometrically extended Cornell
confinement potential approach scattering states
of the Coulomb-like  potential in ordinary flat space. 
Stated differently, the TEC confinement gradually fades away with vanishing
curvature and allows for deconfinement.
\end{quote}
The latter is most easily demonstrated for the case of a TEC potential with
a nullified $G$ parameter and reduced to the
$\csc^2$ term, the $S_R^3$ centrifugal barrier.
Indeed, for small curvatures such that $K \sqrt{\kappa }\sim \mbox{k}$, 
with ``$\mbox{k}$'' a constant, eq.~(\ref{energy_O(4)}) goes into
\begin{equation}
E_K^{(c=0)}(\kappa ) \stackrel{\kappa \to 0}{\longrightarrow }
\frac{\hbar^2}{2\mu }
{\mbox{k}^2},
\label{sct_st}
\end{equation}
and describes a continuous energy spectrum.
Moreover, 
in parallel with the asymptotic behavior of the energy spectrum 
in eq.~(\ref{sct_st}), also the wave functions from the confinement phase,
${\mathcal S}_{Kl}(\chi ,\kappa )$, approach in same limit
the wave functions relevant for the deconfinement phase which are
the scattering states of the inverse distance potential 
in flat $E_3$ space.
 In order to see this it is useful to
recall the following differential recursive relation 
satisfied by  the ${\mathcal S}_{Kl}$ functions \cite{Vinitski},
\begin{eqnarray}
{\mathcal S}_{Kl}(\chi , \kappa) &=&
\frac{\sin^l \chi}{ \sqrt{(n^2-1)...(n^2-l^2)}}
\frac{\mbox{d}^l}{(\mbox{d}\cos \chi)^l}{\mathcal S}_{K0}(\chi ,\kappa),
\nonumber\\
{\mathcal S}_{K 0}(\chi, \kappa ) &=&
\sqrt{\frac{2\kappa }{\pi}}\frac{\sin (K+1)\chi}{\sin \chi}.
\label{rec_S_KL}
\end{eqnarray}
In the limits when  $\kappa \to 0 $, and $\chi \to 0$ 
 in such a way that $\chi/\sqrt{\kappa } $ stays finite and
approaches $ \chi/\sqrt{\kappa} \to  r$ 
(here, the factor $\pi$ has been absorbed by $r$ for simplicity), while 
$K\sqrt{{\kappa} }\to $k
with a constant ``k'', i.e.,
\begin{equation}
\lim _{\kappa \to 0}(K+1)\sqrt{{\kappa} }\to \mbox{k} ,\quad 
\lim_{{\kappa }\to 0}\frac{\chi}{\sqrt{\kappa }} \to  r ,  
\label{limits}
\end{equation}
one also finds
\begin{eqnarray}
\sin (K+1)\chi &\longrightarrow & 
 \sin (K+1)\sqrt{{\kappa} }r\longrightarrow \sin \mbox{k} r,
\quad 
\sin \chi \to \sqrt{{\kappa} }{r},\nonumber\\
\mbox{d}\cos \chi & =&-\sin\chi \mbox{d}\chi \longrightarrow
 -{\kappa } r \mbox{d}r.
\label{limits_more}
\end{eqnarray}
Accounting for the latter relations,
eq.~(\ref{rec_S_KL}) takes the form 
of the Reyleigh formula \cite{Arf_Web}
for the spherical Bessel functions,
\begin{eqnarray}
\lim_{\kappa  \to 0}
{\mathcal S}_{Kl}(\chi ,\kappa )&\to&
\sqrt{\frac{2 \mbox{k}^2}{\pi}}
(-1)^l (\mbox{k} r)^l\left( \frac{1}{\mbox{k} r}\frac{\mbox{d}}
{\mbox{d} (\mbox{k} r)}\right)^l\frac{\sin\mbox{k} r}{\mbox{k} r }\nonumber\\
&=&\sqrt{\frac{2\mbox{k}^2}{\pi}}j_l( \mbox{k}r).
\label{limit_wafu_1}
\end{eqnarray}
The latter wave functions are precisely the ones that describe  scattering
states in ordinary flat $E_3$  space, i.e., 
they are the radial functions in the Helmholtz equation
describing free motion in $E_3$.
This example, though a very simplistic  one, 
is already illustrative of the effect that the curvature shut-down
can have as  deconfinement mechanism.

In the presence of the $\cot \chi $ barrier the spectrum is shaped after
eq.~(\ref{enrg_cot}). In the unconditional $\kappa \to 0$ limit,
the second term of the r.h.s. vanishes and the spectrum 
becomes the one of $H$ atom-like  bound states.
In the conditional $\sqrt{\kappa }K\to \mbox{k}$ limit from above,
where ``k'' is a constant, the term in question approaches the scattering
continuum. In effect, the ${\mathcal V}(\chi ,\kappa )$ 
spectrum  collapses down to the regular Coulomb-like potential, 
\begin{equation}
E_{K}(\kappa )\stackrel{\kappa \to 0}{\longrightarrow} 
-\frac{G^2}{\frac{\hbar^2}{2\mu}}   \frac{1}{n^2}
+ \frac{\hbar ^2}{2\mu }\mbox{k}^2, \quad l=0,1,2,...,n-1.
\label{bua}
\end{equation}
The rigorous proof that also the wave functions the complete TEC potential
collapse to those of the corresponding Coulomb-like problem for 
vanishing curvature is a bit more 
involved and can be found in \cite{Barut}, \cite{Vinitski}.
\begin{quote}
In other words, as  curvature goes down as it can happen because
of its thermal dependence, 
confinement fades away, an observation that is suggestive of
a deconfinement scenario controlled by the curvature parameter
of the TEC potential.
\end{quote}
Deconfinement as gradual flattening of space has earlier been considered
by Takagi \cite{Takagi}.  Compared to \cite{Takagi}, 
our scheme brings the advantage that 
the flattening of space is paralleled  by 
a temperature evolution of the curved TEC-- to a flat
Coulomb-like potential, and correspondingly,
by the  temperature evolution of the TEC wave functions 
from the confined to the Coloumb-like wave-functions from
the deconfined phases, in accordance 
with eqs.~(\ref{limit_wafu_1}), and (\ref{bua}).

\section{Summary}

We emphasized importance of designing confinement phenomena
in terms of infinite potential barriers emerging on curved spaces.
Especially, quark confinement and QCD dynamics have been
modeled in terms of a trigonometric potential that emerges as
harmonic potential on the three-dimensional hypersphere of constant
curvature, i.e., a potential that satisfies the Laplace-Beltrami 
equation there. The potential under consideration 
interpolates between 
the $1/r$-- and infinite well potentials while passing through a region
of linear growth. This trigonometric confinement potential  is
exactly solvable at the level of the Schr\"odinger equation and
moreover, contains the Cornell potential predicted by 
Lattice QCD and topological field theory 
\cite{tHooft},\cite{Witten}, \cite{Hugo}
as leading terms of its Taylor decomposition.
When employed as a quark-diquark potential, it led to a remarkably 
adequate description of the $N$ and $\Delta $ spectra in explaining their
$O(4)/SO(2,1)$ degeneracy patterns, level splittings, 
number of states, and proton electric charge-form factor.
Moreover, the trigonometrically extended Cornell (TEC) potential,
in carrying simultaneously the $SO(4)$ and $SO(2,1)$ symmetries
(as the $H$ atom!),  matches the algebraic aspects of
$AdS_5/CFT$ correspondence and establishes its link to QCD potentiology.
A further advantage of the TEC potential is
the possibility to employ  its curvature parameter, 
considered as temperature dependent,  as a driver of the 
confinement-deconfinement transition in which case the wave functions of
the confined phase approach bound and scattering states of ordinary flat space
$1/r$ potential.

All in all, we view the concept of curved spaces as a promising one
especially within the context of quark-gluon dynamics.

\vspace{0.2cm}

\noindent
{\bf Acknowledgments}\\

\noindent
One of us (M.K) acknowledges hospitality by the Argonne National Laboratory
in April 2008 and stimulating interest and discussions with T.S.H. Lee. 
We thank M. Krivoruchenko 
for providing assess to ref.~\cite{Ismst} and Nora Breton for assistance
in clarifying some aspects of $AdS_5$ gravity.
Work supported by CONACyT-M\'{e}xico under grant number
CB-2006-01/61286.

\end{document}